\documentclass[%
 reprint,
 superscriptaddress,
 nofootinbib,
 amsmath,amssymb,
 aps,
prab,
]{revtex4-2}

\usepackage{graphicx}
\usepackage{dcolumn}
\usepackage{bm}
\usepackage{hyperref}

\usepackage{siunitx}
\usepackage{subcaption}
\usepackage{newtxtext}
\usepackage{tabularx}

\begin{document}

\preprint{APS/123-QED}

\title{Simulation Study of the Space Charge Limit in Heavy-ion Synchrotrons}

\author{Adrian Oeftiger}
\affiliation{GSI Helmholtzzentrum  f\"{u}r Schwerionenforschung GmbH, Planckstr. 1, 64291 Darmstadt, Germany}
\author{Oliver Boine-Frankenheim}%
\affiliation{GSI Helmholtzzentrum  f\"{u}r Schwerionenforschung GmbH, Planckstr. 1, 64291 Darmstadt, Germany}
\affiliation{Technische Universit\"{a}t Darmstadt, Schlossgartenstr. 8, 64289 Darmstadt, Germany}
\author{Vera Chetvertkova}
\affiliation{GSI Helmholtzzentrum  f\"{u}r Schwerionenforschung GmbH, Planckstr. 1, 64291 Darmstadt, Germany}
\author{Vladimir Kornilov}
\affiliation{GSI Helmholtzzentrum  f\"{u}r Schwerionenforschung GmbH, Planckstr. 1, 64291 Darmstadt, Germany}
\author{Dmitrii Rabusov}
\affiliation{Technische Universit\"{a}t Darmstadt, Schlossgartenstr. 8, 64289 Darmstadt, Germany}
\author{Stefan Sorge}
\affiliation{GSI Helmholtzzentrum  f\"{u}r Schwerionenforschung GmbH, Planckstr. 1, 64291 Darmstadt, Germany}

\date{\today}

\begin{abstract}
The SIS100 synchrotron as a part of the new FAIR accelerator facility at GSI should be operated at the ``space charge limit'' for light- and heavy-ion beams.
Beam losses due to space-charge-induced resonance crossing should not exceed a few percent during a full cycle. 
The recent advances in the performance of particle tracking tools with self-consistent solvers for the 3D space charge forces now allow us to reliably identify low-loss areas in tune space, considering the full \SI{1}{\second} (160'000 turns) accumulation plateau in SIS100.
A realistic magnet error model, extracted from precise bench measurements of the SIS100 main dipole and quadrupole magnets, is included in the simulations.
Previously such beam dynamics simulations required non-self-consistent space charge models. 
By comparing to the self-consistent simulations results we are now able to demonstrate that the predictions from such faster space charge models can be used to identify low-loss regions with sufficient accuracy. 

The findings are applied by identifying a low-loss working point region in SIS100 for the design FAIR beam parameters. The bunch intensity at the space charge limit is determined. Several counter-measures to space charge are proposed to enlarge the low-loss area and to further increase the space charge limit. 
\end{abstract}

\maketitle


\section{Introduction}

The SIS100 heavy-ion synchrotron \cite{Spiller2020a} is designed for the production of high-intensity heavy-ion and proton beams at medium energies.
The SIS100 with a circumference of \SI{1080}{\meter} and \SI{100}{\tesla\meter} rigidity will be the main synchrotron of the new Facility for Antiproton and Ion Research (FAIR). 
Key components of the synchrotron are the fast-ramping superconducting magnets (with up to \SI{4}{\tesla/\second}), enabling short cycles, as well as dedicated RF cavities for bunch compression, enabling the extraction of a single short (\SI{50}{\nano\second}) ion bunch at the end of the cycle. 

The magnet systems are currently in production, with all the dipole magnets already tested successfully, showing the required high precision in field quality.
The availability of a full SIS100 magnet error model with nonlinear field distortions measured up to 10${}^\text{th}$ order enables realistic predictions of the SIS100 performance with space charge. 
First results based on a transversely linear space charge model have been reported in Sec.\ 6 of Ref.~\citealp{jinst-sis100-losses}.
Series production of the quadrupole magnets is currently ongoing \cite{borisov:ipac2021-tupab383} with more than fifteen percent having been tested already, which warrants a realistic prediction for the full series.

The main beam loss mechanism expected for intense heavy-ion beams in SIS100 at injection energy -- besides charge stripping of such intermediate charge state ions -- is space-charge-induced resonance crossing.
For proton and light-ion beams space charge is expected to be the main intensity limitation at injection energy and for bunch compression \cite{yuan2018} at top energy. 
In the case of (intermediate charge state) heavy-ions, the tolerance for space-charge-induced beam loss is determined by the effectiveness of the vacuum system to stabilise the residual gas pressure against desorption due to the lost particles.
In addition, and for all ions, beam loss has to remain below certain threshold values to avoid activation and subsequent component damage. 
A beam loss budget of $5\%$ for the injection plateau and intermediate-charge-state heavy ions is considered as tolerable \cite{Spiller2020a}.
Dedicated collimation systems for absorbing `halo' losses caused by stripping and space charge are foreseen in SIS100 \cite{Strasik2015}. 

The six arcs in SIS100 are optimised for the collimation of stripping losses in heavy-ion beams.
The straight sections in between are used for the rf cavities and injection/extraction as well as laser cooling of selected ion beams \cite{Eidam2018}. 
For the generation and subsequent storage of secondary beams downstream, the SIS100 should accelerate primary ion beams from protons up to uranium. 
In our studies we focus on the heavy-ion design intensities and cycles, with the goal to determine the space charge limit and identify knobs to increase it.

The bunches are injected from the existing SIS18 synchrotron \cite{sis18,BOINEFRANKENHEIM2018122} at a maximum repetition rate of \SI{2.7}{\hertz} \cite{Spiller:2014qqa}.
In the SIS100 space charge effects will play an important role especially during the long (approx \SI{1}{\second}) accumulation plateau at the lowest (injection) energy, which is \SI{200}{\mega\electronvolt/u} for U${}^{28+}$ ions, chosen as a reference case in this study. 
After acceleration (up to \SI{2.7}{\giga\electronvolt/u} for  U${}^{28+}$) the beam is either extracted slowly over one or more seconds or re-bunched and compressed before fast extraction.

At injection energy, heavy-ion beams fill up to approximately half of the transverse size of the SIS100 vacuum tube aperture. 
Besides direct space charge forces also image forces from the elliptical beam pipe can modify the incoherent tune shifts and will be included in future simulation models.
In the present contribution we focus on the effect of direct space charge and neglect the wall impedance.

The beam dynamics simulations have to resolve the nonlinear 3D space charge forces as well as the local magnet errors, which are represented by a multipole expansion. 
The coefficients of the multipole are extracted from magnetic field measurements of individual magnets.
Until recently, predictions of space-charge-induced beam loss due to the crossing of resonances have relied on particle tracking using 3D frozen space charge models (see e.g.\ Refs.~\citealp{Asvesta2018,Bartosik2020}).
Due to the demanding time scales of up to 1 second or 160'000 turns, non-self-consistent space charge models are also employed for the 2D tune scans for SIS100, leading to the identification of suitable low-loss working point areas. 
With non-self-consistent frozen space charge models the simulations can be run with a relatively low number of simulation particles. 
Frozen space charge models, also in their adaptive forms \cite{Bartosik2016}, suppress all modifications of the space charge force resulting from non-Gaussian profiles. 
In the non-adaptive (or ``fixed'' frozen) form, also coherent beam oscillations of any sort are entirely suppressed.
It is therefore of vital interest to carefully validate the low-loss areas obtained with the faster frozen space charge models by comparison with self-consistent PIC computation.
Over the long accumulation time scales in synchrotrons, self-consistent PIC solvers require large numbers of macro-particles in order to limit the effect of artificial noise \cite{Boine-Frankenheim2014, Hofmann2014a}. 
Recent implementations of the self-consistent Particle-In-Cell (PIC) scheme feature a much improved performance with hardware acceleration \cite{oeftiger2016space}. 
We are hence now in the position to use a sufficiently large number of macro-particles for effective noise suppression and validate the identified low-loss areas, within acceptable computing times.
This, for the first time, enables systematic studies of different optimisation measures in a realistic simulation scenario towards an increase of the space charge limits.

The first goal of our studies is to identify suitable working point areas for low-loss operation in view of the high nominal bunch charge and given the SIS100 magnet error model, based on the magnet bench measurements.
The second goal is to identify suitable optimisation measures, to further widen the low-loss areas for the design beam intensity, in order to allow for more options in tune space during high-intensity operation. Thirdly we define and explore the space charge limit based on our realistic computer model.

This contribution is organised as follows: first we discuss the SIS100 design beam parameters and our 3D simulation model. 
We proceed to analyse the ideal, error-free SIS100 lattice with only space charge included.
Next, the quadrupole resonance driven by a gradient error serves as a study case to compare and establish predictions by the space charge models.
In the central part we include the full nonlinear magnet error model and focus on beam loss prediction and optimisation.
We conclude with a summary and outlook.

\section{The SIS100 Beam-loss Simulation Model}

The particle transport through the accelerator lattice is computed employing the 6D symplectic single-particle tracking engine library SixTrackLib \cite{sixtracklib}.
SixTrackLib is optimised for running on highly parallel hardware architectures, as such it supports running simulations on multi-core CPUs as well as graphical processing units (GPUs).

Our simulation model includes a detailed SIS100 lattice consisting of 84 basic focusing cells, which are arranged in six symmetric arcs and straight sections.
The symplectic tracking algorithm of SixTrackLib necessitates a thin-lens representation of the extended magnets\footnote{The dipole and quadrupole units are represented by 9 thin kicks which are located by using the TEAPOT algorithm \cite{teapot}.}.
The second-order fringe field effect of the dipole magnets is taken into account in hard-edge approximation\footnote{The additional focusing contribution by the dipole magnet fringe fields makes for a $0.1$ tune shift in the SIS100 lattice.}.

For machine protection in real operation, SIS100 features a collimation system as well as a cryo-catcher system with low desorption surfaces to avoid dynamic vacuum effects \cite{cryocatchers}. 
By construction they essentially provide the most stringent aperture limitation in order to absorb halo particles and ions of non-design charge state.
To simplify the simulation setup for this study, the collimators and cryocatchers are retracted in the machine model and thus not taken into account.
The loss model hence covers the apertures of the magnets as well as the extraction septa of the machine. 

We focus on the SIS100 reference U${}^{28+}$ heavy-ion beam at injection energy, for which Table \ref{tab: params} lists the machine and beam parameters. 
Also lighter ions and protons with similar space charge tune shifts will be accelerated. 
However, U${}^{28+}$ beams feature the lowest injection energy. 
The correspondingly large beam sizes compared to the physical aperture render beam loss mechanisms induced by space charge particularly important: any significant emittance growth translates to finite beam loss.
In addition the loss budget implied by vacuum considerations is the tightest for U${}^{28+}$ beams. 
Therefore this case serves as the reference throughout the rest of the paper.

\begin{table}[htbp]
    \centering
    \caption{Considered parameters for ${}^{238}$U${}^{28+}$ accumulation at SIS100 injection energy.}
    \renewcommand{\arraystretch}{1.3}
    \label{tab: params}
    \begin{tabular}{r|l}
        \hline\hline
        \textbf{Parameter} & \textbf{Value} \\ \hline
        Horizontal normalised rms emittance\footnote{The normalised rms emittance corresponds to a ``Kapchinskij-Vladimirskij (KV)'' emittance (as quoted e.g.\ in Ref.~\citealp{Franchetti2006b}) of \SI{35}{\milli\meter.mrad} and \SI{15}{\milli\meter.mrad}, which is 4 times the geometric RMS emittance.} $\epsilon_x$ & \SI{5.9}{\milli\meter.mrad} \\
        Vertical normalised rms emittance\footnotemark[1] $\epsilon_y$ & \SI{2.5}{\milli\meter.mrad} \\
        Rms bunch length $\sigma_z$ & \SI{13.2}{\meter} \\
        Rms momentum deviation $\sigma_{\Delta p/p_0}$ & \SI{0.44d-3}{} \\
        Bunch intensity $N$ of U${}^{28+}_{238}$ ions & \SI{0.625d11}{} \\
        Max.\ space charge tune shift $\Delta Q^\text{SC}_y$ & $-0.30$ \\
        Chromaticity $Q'_{x,y}$ & $(-21, -24) $ \\
        RF voltage (single-harmonic) $V_\text{RF}$ & \SI{58.2}{\kilo\volt} \\
        Harmonic $h$ & $10$ \\
        Transition energy $\gamma_{t}$ & $15.5$ \\
        Synchrotron tune $Q_s$ & \SI{4.5d-3}{} \\
        Circumference $C$ & \SI{1083.6}{m} \\
        Kinetic energy & $E_{kin}=\SI{200}{\mega\electronvolt}/$u \\
        Relativistic $\beta$ factor & 0.568 \\
        Revolution frequency $f_\text{rev}$ & \SI{157}{\kilo\hertz} \\ \hline\hline
    \end{tabular}
\end{table}

Coming from the upstream SIS18 synchrotron, the transferred bunches are foreseen to be scraped in the transfer line. 

\begin{figure}[htb]
    \centering
    \includegraphics[width=\linewidth]{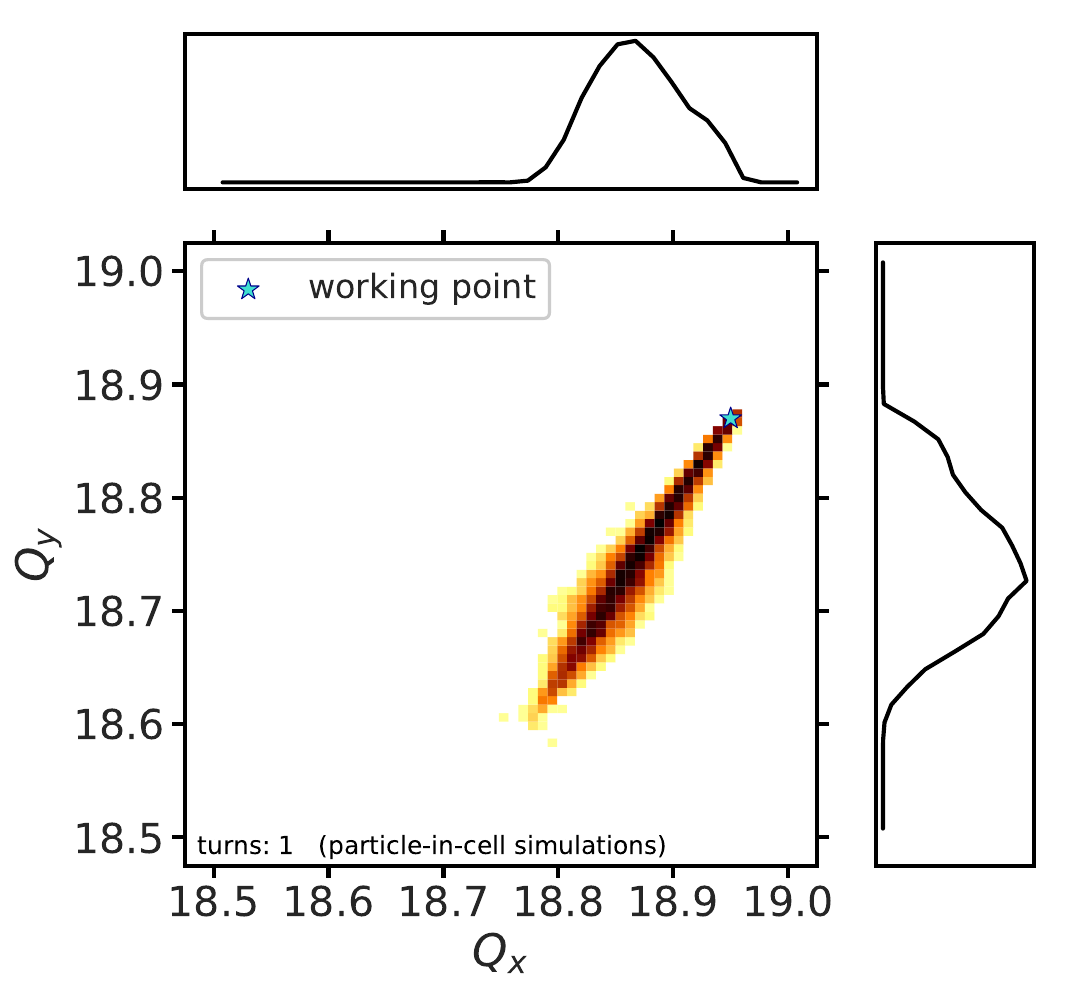}
    \caption{Incoherent tune footprint of a SIS100 U${}^{28+}$ bunch at injection with space charge (using the 2.5D PIC space charge model) and natural chromaticity as tune spread sources.}
    \label{fig:tunespread}
\end{figure}

The simulations for SIS100 injection take the scraping effect into account by generating 6D Gaussian bunch distributions which are cut at $2\sigma$ (with resulting rms figures according to Table \ref{tab: params}).
The incoherent footprint of such a bunch distribution subject to space charge and natural chromaticity is plotted in Fig.~\ref{fig:tunespread}.
The (defocusing) space charge effect with a maximum detuning of $\Delta Q_y^\text{SC}=-0.3$ in the bunch centre dominates over the (linear) chromatic tune spread of $Q'_{x,y}\sigma_{\Delta p/p_0}=0.01$. 
Both effects are included in the simulations.

For reference, the maximum transverse detuning by space charge in a Gaussian distributed bunch can be determined analytically \cite[Eq.~(2.87)]{oeftigerthesis} by
\begin{align} \label{eq: sc tune shift}
    \begin{split}
    \Delta Q_y^\text{SC} = - &\frac{Ze}{4\pi\epsilon_0 m_0 c^2}\,\frac{\lambda_{max}}{\beta^2\gamma^3} \times \\
    \times &\frac{1}{2\pi}\oint ds \frac{\beta_y(s)}{\sigma_y(s) \bigl(\sigma_x(s)+\sigma_y(s)\bigr)} \quad ,
    \end{split}
\end{align}
where $Z$ is the ion charge number, $e$ the elementary charge, $\epsilon_0$ the vacuum permittivity, $m_0$ the rest mass of the ions, $c$ the speed of light, $\beta$ the speed in units of $c$, $\gamma$ the Lorentz factor, $\beta_y(s)$ the vertical betatron function along the path length $s$ around the accelerator, and $\sigma_{x,y}$ the horizontal or vertical rms beam size.
The maximum line charge density $\lambda$ for a Gaussian shaped bunch of rms bunch length $\sigma_z$ and intensity $N$ reads
\begin{equation}
    \lambda_{max} = \frac{NZe}{\sqrt{2\pi}\sigma_z} \quad .
\end{equation}
It is interesting to note that, due to the particular SIS100 lattice layout which is optimised for the catching efficiency of stripped ions, the contribution of dispersion to the beam size $\sigma_x$ is negligibly small when computing the tune shift in Eq.~\eqref{eq: sc tune shift}.

\subsection{Space charge models and solvers}

In our study we use two different approaches to account for the 3D space charge force. 
The models are implemented as lumped kicks in the PyHEADTAIL tracking code \cite{Oeftiger:2672381,oeftiger2016space} which is integrated with SixTrackLib for the SIS100 simulation model. 
In all approaches we only account for the direct space charge force and neglect indirect wall effects. 

In the fast and approximative approach, the transverse space charge force is obtained from a frozen Gaussian distribution using the expression derived in Ref.~\cite{bassetti}. 
In the longitudinal plane, a Gaussian line charge density shape is assumed.
Considering a static force field map without updates throughout the simulation, this approach will be henceforth referred to as Fixed Frozen Space Charge (FFSC). 
The \emph{Adaptive} Frozen Space Charge model (AFSC) then takes into account the change of rms beam size of the evolving simulated macro-particle distribution while still assuming a Gaussian space charge force. 
If the field map update only takes place every few turns, also in the AFSC model coherent first- and second-order beam oscillations are suppressed.

The 2.5D PIC employed throughout this study solves the transverse Poisson equation for many longitudinal slices along the bunch\footnote{The full 3D PIC solver does not yield significantly different results compared to the slice-by-slice 2.5D PIC solver for the elongated bunches in SIS100, which justifies the use of the faster 2.5D algorithm. The grid spans 128 cells per transverse plane and 64 slices along the longitudinal plane, where the bunch is resolved with transversely at least 5 cells per rms beam width and longitudinally more than 14 slices per rms bunch length.}.
Over the long accumulation time scales in synchrotrons PIC requires large macro-particles numbers\footnote{Given the above grid dimensions and a macro-particle number of 10 million, there are thus on average several thousand macro-particles per cell at peak current which appears to sufficiently resolve the bunch dynamics in the present long-term study.} in order to limit the effect of artificial noise \cite{Boine-Frankenheim2014, Hofmann2014a}. 
In the following sections we will compare results obtained with FFSC (and AFSC where appropriate) to PIC over extended parameter ranges.

\section{Resonances from Space Charge}
\label{sec:bare machine}

In this section we apply our simulation model to the ideal SIS100 lattice, with space charge as the only source for nonlinear resonance driving terms. 
The SIS100 lattice consists of superconducting superferric dipole and quadrupole magnet units \cite{mierau2016testing}, with the exception of two radiation hardened and thus normal-conducting quadrupole magnets in the extraction section. 
In Section~\ref{sec: warm lattice} we will analyse the perturbation given by these two ``warm'' quadrupole magnets. 
Otherwise we will consider the perturbation caused by the two warm quadrupole magnets to be compensated, yielding a ring super-periodicity of $S=6$ in the ideal SIS100 layout. 
We focus on the tune quadrant $18.5<Q_{x,y}<19$, which has been identified as a possible candidate for operation with heavy ions and fast extraction \cite{Franchetti2006b}. 

From the optics design point of view, one should avoid tune quadrants with structure resonances in order to push the space charge limit in synchrotrons, especially if long-term storage during an accumulation plateau is required.
Structure resonances driven by space charge can be categorised into mainly two types, viz.\ coherent instabilities and incoherent resonances due to periodic modulation of the space charge potential.
The former, instabilities of coherent modes, have been shown in Ref.~\cite{PhysRevAccelBeams.24.024201} to be Landau damped in Gaussian beams for nonlinear orders -- only the envelope (second-order) instability remains.
After the duration of a few synchrotron periods the envelope instability stop-band is entirely surrounded by the incoherent fourth-order resonance stop-band.
Therefore, coherent phenomena are expected to be either absent or at least covered within the stop-bands reproduced by incoherent frozen space charge models.

Following this argumentation, in this paper the FFSC model will serve to predict the extent of both internally (space charge) and externally (lattice / error) driven stop-bands.
The predicted results will be confirmed and validated with extensive self-consistent PIC simulations for the duration of $\sim 100$ synchrotron periods.

In absence of magnet field errors, beam loss mainly originates from space-charge-induced incoherent resonances.
As illustrated in Ref.~\cite{absence-structure-resonances-sis100}, the design tune quadrant for heavy-ion beam production is clear of structure-resonance-induced beam loss up to order $n=6$.
Structure resonances of lower than 6${}^\text{th}$ order outside the design working point area are observed to induce beam loss in SIS100.
On the other hand, the CERN PS has been shown to suffer from an 8${}^\text{th}$ order (basic focusing period) structure resonance during the injection plateau \cite{PhysRevAccelBeams.23.091001}.
All in all, the usual incoherent resonance diagrams with structure resonances plotted up to at least $n=8$ should thus provide a good picture which tune areas are to be avoided, where resonances may affect Gaussian bunches stored during long injection plateaus in synchrotrons.

The existing SIS100 correction magnets are not used in the present study, therefore the bare machine provides no further driving terms for externally driven resonances.
In the absence of magnet errors there is only the space charge driven Montague difference resonance \cite{montague1968fourth} present in the tune quadrant, located around the $Q_x=Q_y$ coupling line.
In order to compare the different space charge modelling approaches and their limitations, we will first focus on this simple but relevant example case.



\subsection{Montague Difference Resonance}
\label{sec: montague}

Transversely Gaussian distributed beams are subject to the well studied Montague fourth-order difference resonance mechanism \cite{montague1968fourth,PhysRevSTAB.9.054202}:
the skew octupole component in the space charge potential drives a transverse emittance exchange around the coupling line according to the resonance condition
\begin{equation} \label{eq: montague}
    2Q_x-2Q_y = 0 \quad .
\end{equation}

The SIS100 beams feature a transverse emittance ratio of more than two.
Therefore, choosing a working point within the Montague stop-band leads to an effective action transfer from the horizontal to the vertical plane.
The vertical emittance grows and particles gaining a large amplitude can get lost in the smaller SIS100 machine aperture in the vertical plane.

We simulate the emittance exchange around the coupling resonance with PIC and the frozen models.
Fixing the horizontal tune at $Q_x=18.75$, the vertical tune is scanned across the resonance condition Eq.~\eqref{eq: montague} which is met at $Q_y=Q_x=18.75$.
The simulations are based on the symmetric lattice, i.e.\ the gradient error from the two warm quadrupole magnets is absent.

\begin{figure}[htbp]
    \includegraphics[width=\linewidth]{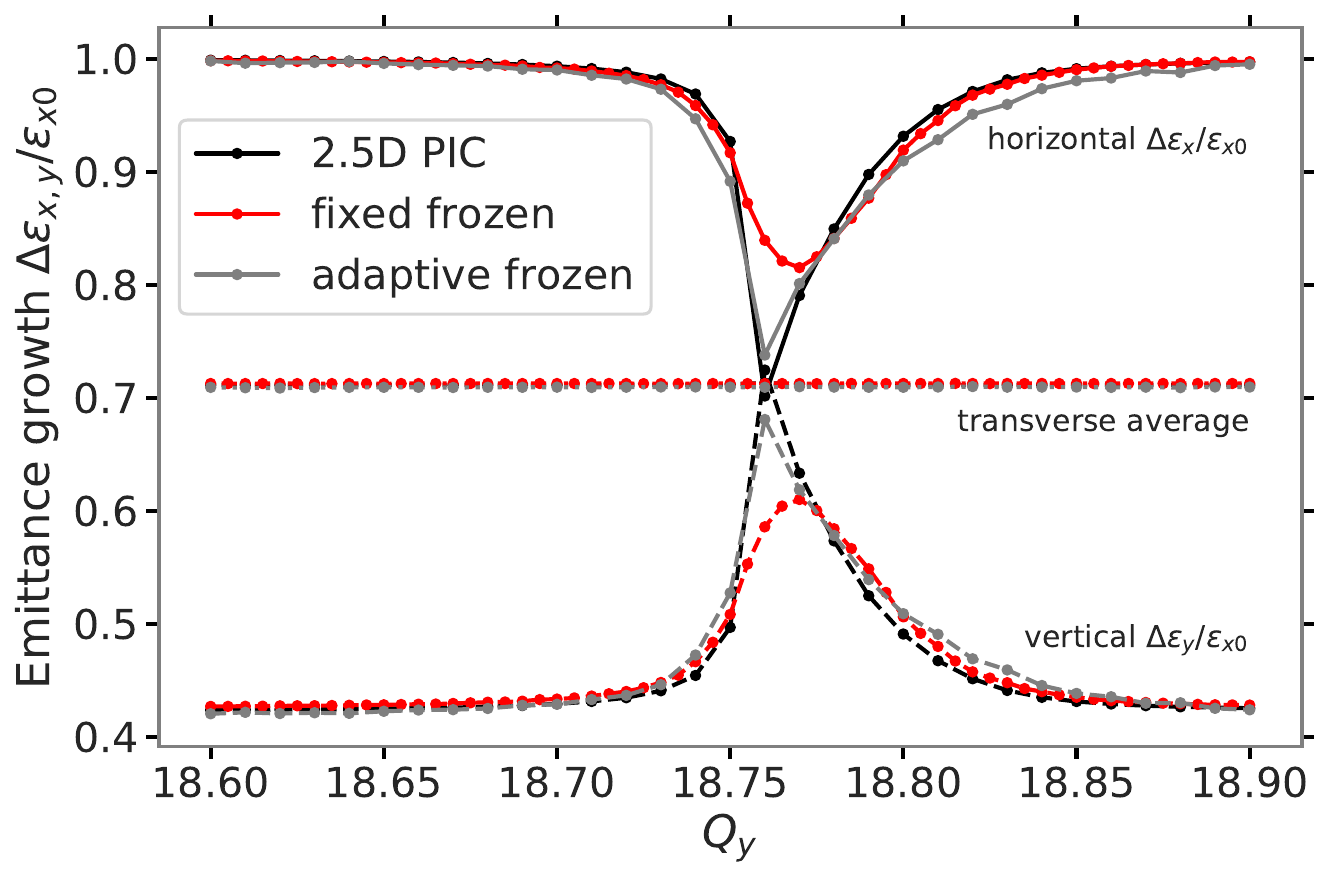}
    \caption{Simulated emittance exchange across the Montague resonance for different space charge models. Solid lines (top part) indicate the horizontal plane, dashed lines (bottom part) the vertical plane and the constant dash-dotted lines (centre) slightly above 0.7 the transverse average.}
    \label{fig:montague}
\end{figure}

The simulation results after saturation\footnote{200 turns are sufficient to capture the full dynamics of the Montague resonance, the emittance curves reach a saturated state of equilibrium.} are presented in Fig.~\ref{fig:montague} for all three space charge variants.
The final horizontal emittances predicted by each variant are marked by the three solid lines, the vertical emittances by the three dashed lines.
Both emittances are given in units of initial horizontal emittance, therefore the solid line starts at 1.0 (no horizontal emittance decrease) and the dashed line at 0.45 (corresponding to the initial ratio between vertical and horizontal emittance).
The dash-dotted line above 0.7 marks the final averaged transverse emittance: since it is constant across all scanned tunes, there is indeed only emittance exchange between both planes and no additional emittance growth due to quadrupole or other resonance stop-bands.
The Montague stop-band as predicted by the self-consistent PIC model extends from about $Q_y\approx 18.7$ ($\Delta Q_y=-0.05$ below the resonance condition) to about $Q_y\approx 18.85$ ($\Delta Q_y=0.15$ above).
This asymmetry finds its origin in the aspect ratio of the transverse space charge tune spread, $|\Delta Q^\text{SC}_x|=0.2 < 0.3 = |\Delta Q^\text{SC}_y|$.

The PIC results (in black) demonstrate that the emittance exchange leads to a balance of the final transverse emittances around the resonance condition $2Q_x-2Q_y=0$.
At $Q_y=18.76$, slightly above the resonance condition, one even observes a small overshoot with final $\epsilon_y>\epsilon_x$.
The FFSC results (in red) apparently fail to resolve this full exchange, the natural reason behind is the static nature of the space charge fields.
Nevertheless, the PIC computed extent of the Montague stop-band is predicted quite accurately by the FFSC model.

The AFSC model (in grey), with updates of the rms envelopes every turn, recovers the PIC results quite accurately w.r.t.\ the full emittance exchange around $Q_x\approx Q_y$.
At the same time, the overall stop-band width appears wider for AFSC.

Our study aims to identify low-loss working points with fast and approximative space charge models.
Therefore, predicting the correct stop-band extent is more important than better resolving the full emittance exchange.
We conclude that FFSC (as opposed to its adaptive variant) covers the extent of the Montague stop-band with satisfying precision, such that resonance-free working points near the coupling line can be well reproduced compared to the PIC model.

\section{Error Resonances in the Broken-symmetry Lattice}
\label{sec: warm lattice}

Let us now turn our attention to the actual machine layout, where two superconducting, ``cold'' quadrupole magnets in the extraction region are replaced by normal-conducting, ``warm'' counterparts.
These are required to sustain inevitable beam loss expected during slow extraction of the beam.
As normal-conducting magnets they are operated at room temperature which prevents quenching issues if they heat up due to particle loss.
In comparison to the cold magnets the warm magnets reach a lower local field gradient.
To supply the same integral focusing strength, the warm magnets are designed with an increased length of \SI{1.76}{\meter} which is more than \SI{0.5}{\meter} longer compared to their shorter cold counterparts.
This lattice configuration with warm quadrupole magnets therefore becomes subject to $\beta$-beating, and the sixfold symmetry of the all-cold quadrupole magnets ring layout is broken.

The two warm quadrupole magnets are located next to each other in one of the six straight sections.
The $\beta$-beat can be reduced by choosing an optimal ratio between their individual currents and the general (cold) quadrupole families. By additionally powering two adjacent corrector quadrupole magnets enclosing the warm quadrupole magnets, the  $\beta$-beat can be further reduced \cite{davidtwocorr}.
The standard deviation of the warm-quadrupole-induced $\beta$-beat around an otherwise ideal machine can thus be considerably brought down, reaching a residual level of less than $2\%$. 
The influence of space charge on this correction set-up deserves a thorough analysis and will be subject of a separate publication.

\begin{figure}[htb]
    \centering
    \includegraphics[width=\linewidth]{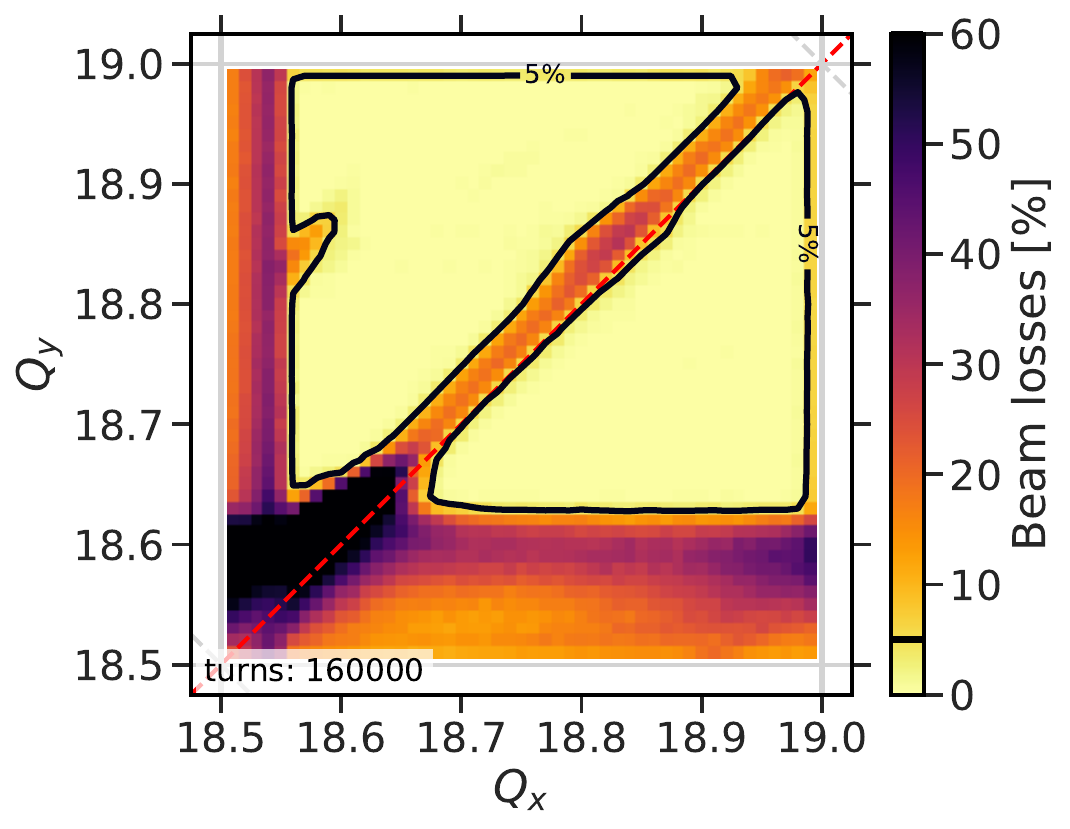}
    \caption{Corrected warm quadrupole magnets. Beam loss tune diagram with FFSC simulation results.}
    \label{fig: warm corrected lattice}
\end{figure}

Here it may suffice to demonstrate that mainly the quadrupole stop-bands remain in the corrected warm quadrupole scenario (besides the Montague resonance).
Beam loss results of FFSC simulations for the $\beta$-beat minimised set-up are depicted in the tune diagram in Fig.~\ref{fig: warm corrected lattice}.
For each simulated working point the colour represents the predicted amount of particles lost across the injection plateau of \SI{1}{\second} or 160'000 turns length, ranging from low loss figures in yellow to intolerably high figures in dark violet.
The black contour encloses the good tune region with less than $5\%$ beam loss.

In the following we consider the uncompensated broken-symmetry scenario, before the remainder of the paper largely assumes the warm quadrupole perturbation to be perfectly compensated by using the symmetric SIS100 lattice.
For now, the induced quadrupolar resonance serves us to investigate the impact of the space charge model on the affiliated stop-band prediction.

\subsection{Gradient errors with space charge}
\label{sec:quad-resonances}

In the absence of space charge the quadrupole resonance condition $2Q_{x,y}=37$ corresponds to the lower corners $Q_{x,y}=18.5$ of the tune quadrant under consideration. 
In the example case studied in the following, the local gradient error results from the two radiation-hard quadrupole magnets in the extraction region, without compensation. 
Consider the two warm quadrupole magnets to be operated at the same integral strength as the corresponding cold quadrupole family,
\begin{equation}
    (K_1 L)_\text{cold} = (K_1 L)_\text{warm} \quad ,
\end{equation}
i.e.\ their inverse focal length is identical when approximated by a single thin lens.

In comparison to the symmetric lattice, the beam experiences a finite focusing gradient in the protruding end regions of the warm quadrupole units\footnote{Possible fringe field effects in the warm quadrupole units are neglected in our treatment.}, whereas the gradient is weaker in the centre region equivalent to the cold quadrupole length.

The warm quadrupole units thus act as a local source of $\beta$-beat around the SIS100 ring.

Without space charge, the vertical harmonic stop-band integral \cite[Eq.\ (2.118)]{lee2018accelerator}
\begin{equation}
    (F_{y})_n= \frac{1}{2\pi} \int_0^{C} ds\, \beta_{y}(s)\, \Delta k(s)\, \exp\left(-i\, n\,\frac{\phi_{y}(s)}{Q_y}\right)
\end{equation}
yields the tune shift via the non-oscillatory $F_0$ as
\begin{equation}
    \Delta Q_{warm} = \frac{F_0}{2} = 0.012 \quad .
\end{equation}
Here, $\phi_{y}$ denotes the unperturbed betatron phase advance, $\beta_{y}$ the unperturbed betatron function and $\Delta k$ the gradient error at path length $s$ around the machine of circumference $C$.

In absence of space charge, the stop-band width\footnote{The tune distance $\delta Q_{stopband}$ relates to the upper and lower edge of the stop-band, where the maximum local $\beta$-beating around the ring reaches the order of magnitude of the local $\beta$-function: $\max\left[\Delta\beta(s)/\beta(s)\right]\approx 1$.}
of the quadrupolar resonance $2Q_y=37$ is expected to be mainly determined by the corresponding harmonic $F_{37}$,
\begin{equation} \label{eq: stopbandwidth}
    \delta Q_{stopband} = \left| F_{37}\right| = 0.023 \quad .
\end{equation}

\begin{figure}[b]
    \centering
    \includegraphics[width=\linewidth]{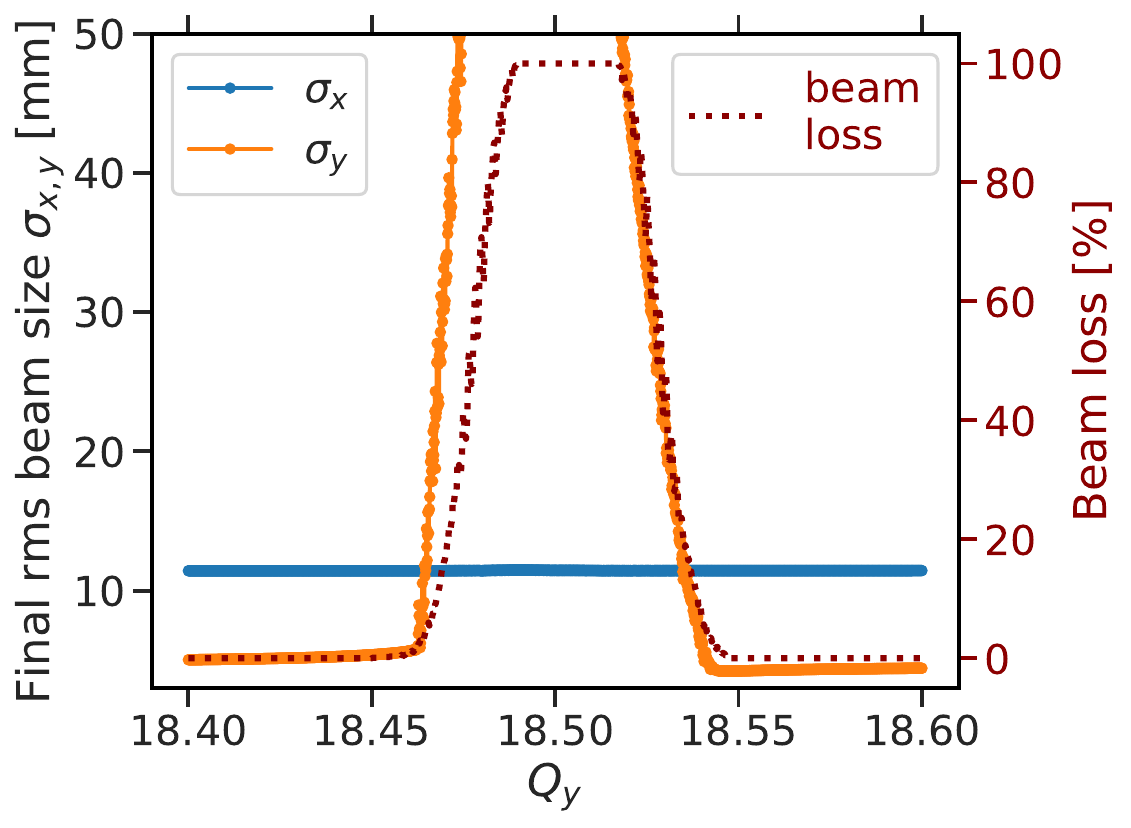}
    \caption{Tracking results without space charge for final rms beam sizes vs.\ vertical tunes around the quadrupole resonance $Q_y=18.5$ at fixed $Q_x=18.75$. Dotted line: beam loss curve for separate simulations including apertures.}
    \label{fig:halfinteger noSC}
\end{figure}

Let us compare this computed stop-band width to tracking simulations without space charge.
Figure \ref{fig:halfinteger noSC} displays the results for fixed horizontal tune $Q_x=18.75$ while varying the vertical tune around the $2Q_y=37$ resonance.

While the horizontal rms beam size (blue) remains constant, the vertical rms beam size (orange) increases rapidly. 
The orange curve exceeds a threshold of $\sqrt{2}\sigma_{y0}$ in the stop-band interval $18.46<Q_y<18.54$ (after simulating two synchrotron periods).

When including apertures in a separate set of simulations, the particles growing in amplitude are marked lost where they reach the aperture.
The result is plotted as the dotted black line in Fig.\ \ref{fig:halfinteger noSC}, indicating finite beam losses between $18.45<Q_y<18.55$. 
The simulated stop-band width of about $\delta Q_{stopband}\approx 0.1$ is larger than the computed value in Eq.~\eqref{eq: stopbandwidth} due to the impact of chromaticity.
For a separate set of simulations with negligibly small momentum spread, the simulated stop-band indeed matches the analytically estimated width.

Both analytically and numerically obtained stop-band widths are smaller than the maximum space charge tune shift $\Delta Q_y^\text{SC}=-0.3$.
With space charge, particles can cross the resonance condition periodically during their synchrotron oscillations. 
The stop-band width is hence expected to be dominated by the space charge tune spread.
The particles at the bunch tails are only weakly affected by space charge and momentarily experience the quadrupole stop-band with nearly no space charge. 
In the bunch centre the space charge tune shift is strongest. 
Here one expects a coherent resonant response to the gradient error due to coherent advantage \cite{fedotov2002half}. 

With the relevant harmonic being 37, the coherent quadrupole resonance condition \cite{baartman1998betatron} reads
\begin{equation} \label{eq: envelope resonance condition}
    2\bigl(Q_y - \mathcal{C}_2 \bigl|\Delta Q_y^{KV}\bigr|\bigr) = 37 \quad ,
\end{equation}
where $\mathcal{C}_2=2/3$ for the vertical plane (using $2\sigma_y=\sigma_x$ in smooth approximation and split tunes, i.e.\ weakly coupled quadrupole modes). 
The rms-equivalent KV tune shift amounts to half the maximum Gaussian tune shift, $2\Delta Q_y^{KV}=\Delta Q_y^\text{SC}$.

\begin{figure}[b]
    \centering
    \includegraphics[width=\linewidth]{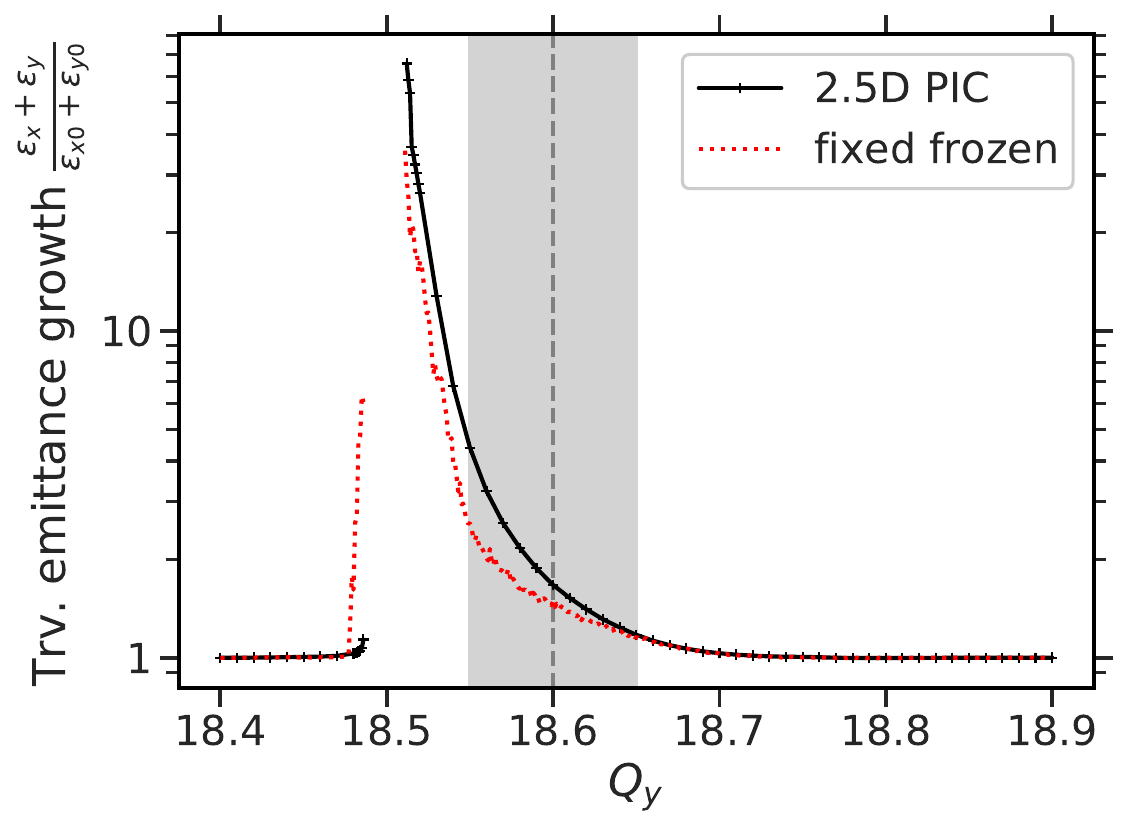}
    \caption{Comparison between PIC and FFSC space charge models. Transverse emittance growth ratio vs.\ vertical tunes around the quadrupole resonance at fixed $Q_x=18.75$. The resonance condition for the bunch centre Eq.~(\ref{eq: envelope resonance condition}) is indicated by the dashed grey line, the grey area around it indicates the zero-intensity stop-band width.}
    \label{fig:halfinteger}
\end{figure}

For the SIS100 reference bunch parameters Fig.~\ref{fig:halfinteger} exhibits the simulation results for the computed transverse emittance growth ratio after 1000 turns around the quadrupole resonance at fixed $Q_x=18.75$.
Apertures and beam losses are not taken into account in these simulations, as we focus on a comparison of the envelope mismatch (relative beam size) generated by the resonance during a fixed simulation time. 
We plot the average transverse emittance to visualise only the gradient-error-induced total emittance growth. 
The averaged transverse emittance is not affected by the present Montague resonance as discussed in the previous section.   
The black curve marks the more realistic self-consistent PIC prediction, while the red curve indicates the FFSC results\footnote{PIC simulations are run with \SI{1d6}{} macro-particles (convergence confirmed with \SI{4d6}{} macro-particles) and FFSC simulations are identical to the zero-intensity case with \SI{1000}{} macro-particles (convergence confirmed with \SI{10000}{} macro-particles).}.

Following the Sacherer treatment \cite[Part I.2.]{sacherer1968transverse}, the upper edge of the quadrupole stop-band can be found via the zero-intensity stop-band shifted by the envelope tune. 
For the bunch centre, the coherent quadrupole resonance condition Eq.~(\ref{eq: envelope resonance condition}) yields $Q_{y}^{coh}=18.6$, which is plotted as a dashed grey line in Fig.~\ref{fig:halfinteger}. 
The filled grey area around it depicts the extent of the zero-intensity stop-band.

As pointed out in the introduction, the constant space charge fields in the FFSC simulation model cannot reproduce coherent phenomena in contrast to the PIC model. 
The modelled interaction of the bunch with the resonance is of purely incoherent nature as the simulated particles experience the resonance without knowing of each other.

We note the following observations: the lower end of the stop-band is shifted upwards for both PIC and FFSC compared to the zero-intensity results from Fig.\ \ref{fig:halfinteger noSC}.
The upper end of the quadrupolar stop-band in terms of finite emittance growth extends until approximately $Q_y = 18.7$ for the space charge cases.

Above $Q_y>18.5$, we further observe FFSC to underestimate the emittance growth in comparison to PIC, within the interval where the emittance growth exceeds $\Delta\epsilon/\epsilon_0>15\%$.
A possible explanation is that PIC takes into account fast coherent resonance mechanisms.
These are expected below the upper edge of the grey area marking $Q_y^{edge,coh.res}=18.64$.
Above this point, both PIC and FFSC curves unite and predict equivalent (but small) emittance growth.

This observation suggests the natural conclusion that the approximative FFSC model can accurately represent the more realistic PIC model in tune areas weakly affected by resonance.
In absence of significant coherent motion, the beam self-fields far away from the core are well represented by the FFSC model.
Thus, the same conclusion applies not only to weak emittance growth ($<15\%$) but also to weak beam loss.
Most important of all, working points free of emittance growth and beam loss will be equally well identified by both FFSC and PIC computations.

These features validate the FFSC model for the search of loss-free working points in extensive tune diagram scans -- a crucial point for the remainder of the present paper.
Analogous to the externally driven resonance here, the same conclusion has been drawn for space charge (internally) driven resonances when comparing FFSC and PIC \cite[Fig.~9]{PhysRevAccelBeams.24.024201}:
the FFSC model proves to be a viable fast prediction tool for ideal working points for machine operation.
Needless to say that this comparison only holds as long as the resonance-induced emittance growth is larger than the inherent noise-related emittance growth in PIC -- an effect which is entirely absent in the FFSC case.

The approximative FFSC model becomes particularly powerful due to the fast computation: (GPU accelerated) PIC simulations require computation times exceeding those of (multi-core CPU accelerated) FFSC simulations by two or even three orders of magnitude\footnote{Simulation times for the full injection plateau of 160'000 turns take 48 days with PIC on the NVIDIA V100 GPU with 20 million macro-particles (cf.\ Fig.~\ref{fig:long-pic-sim}), while FFSC on 16 cores of a recent HPC AMD CPU requires 47 minutes for 1'000 macro-particles (cf.\ Fig.~\ref{fig: tune diagram nonlinSC allerrs}).}.

At this point we emphasise our finding that the \emph{adaptive} frozen space charge model leads to an overly conservative estimate of resonance-free working points compared to the FFSC model while also requiring (at least a factor ten) more macro-particles to suppress artificial noise from the field updates. 
In particular, the upper edge of the stop-band appears to extend further than in both PIC and FFSC cases, cf.~Appendix~\ref{app: afsc}.
For the remainder of the paper we will therefore consider the \emph{fixed} frozen space charge model.

\section{Beam loss due to magnet field errors}
\label{sec: field imperfections}

In this section we will add the nonlinear field errors and provide beam loss predictions with space charge. 
The previous sections dealt with betatron resonances implied by the lattice structure as well as space charge.
Here we will discuss non-structure resonances which are driven by random field imperfections of the dipole and quadrupole magnets.
The stochastic field error model considered in the simulations is based on cold bench measurements and has been described in detail in Ref.~\cite{jinst-sis100-losses}.
A summarising overview is found in Appendix~\ref{app: error model}.

\subsection{Bare Machine with Field Errors}

\begin{figure}[htbp]
    \centering
    \includegraphics[width=\linewidth]{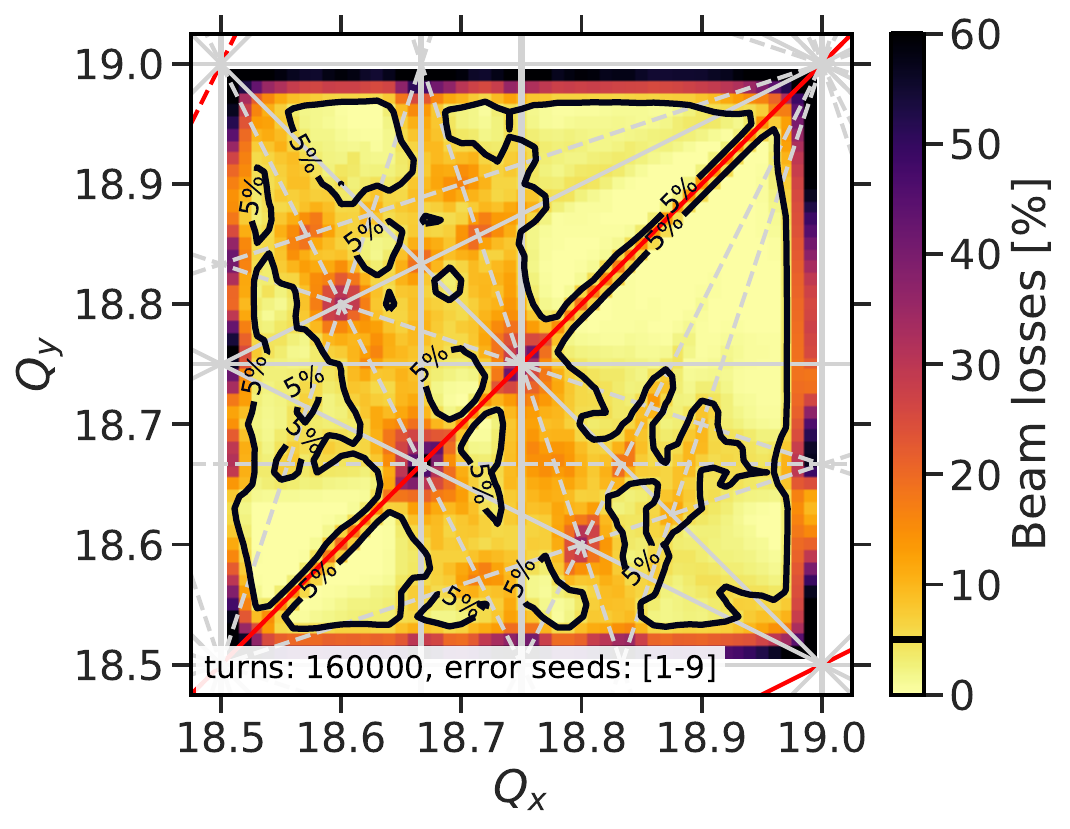}
    \caption{Beam loss tune diagram for vanishing space charge and all magnet field errors included at natural chromaticity.}
    \label{fig: tune diagram noSC allerrs}
\end{figure}

First we will discuss beam loss scans without space charge.
Figure \ref{fig: tune diagram noSC allerrs} shows the resonances in the tune quadrant around $18.5$ to $19$, driven by the magnet field imperfections\footnote{Several seeds for the random distributions have been run across the tune diagram. This is indicated by the text ``error seeds: [1-9]'' in Fig.~\ref{fig: tune diagram noSC allerrs} as well as in all following figures with simulation results including the field error model. While the prediction of finite beam loss varies between different seeds, results agree quite well for working points with vanishing beam loss. The error seed indicated by ``1'' has been found to provide a conservative estimate of low-loss areas among the seeds tested. It is used as a representative in simulation scenarios where only one error sequence has been considered.}.

Several thin lines of significant beam loss can be identified where nonlinear resonances are excited.
The width of the stop-bands is increased by the tune spread due to the natural chromaticity amounting to an rms figure of $0.01$.
The coupling line does not imply strong beam loss, which is in line with the absent space charge driving terms for the Montague resonance.

It is important to point out that, without space charge and only taking into account the realistic magnet error model, the tune quadrant exhibits large areas for low-loss operation.   

\subsection{Including Space Charge}
\label{sec:mainU28}

\begin{figure}[htbp]
    \centering
    \begin{subfigure}{\linewidth}
        \includegraphics[width=\linewidth]{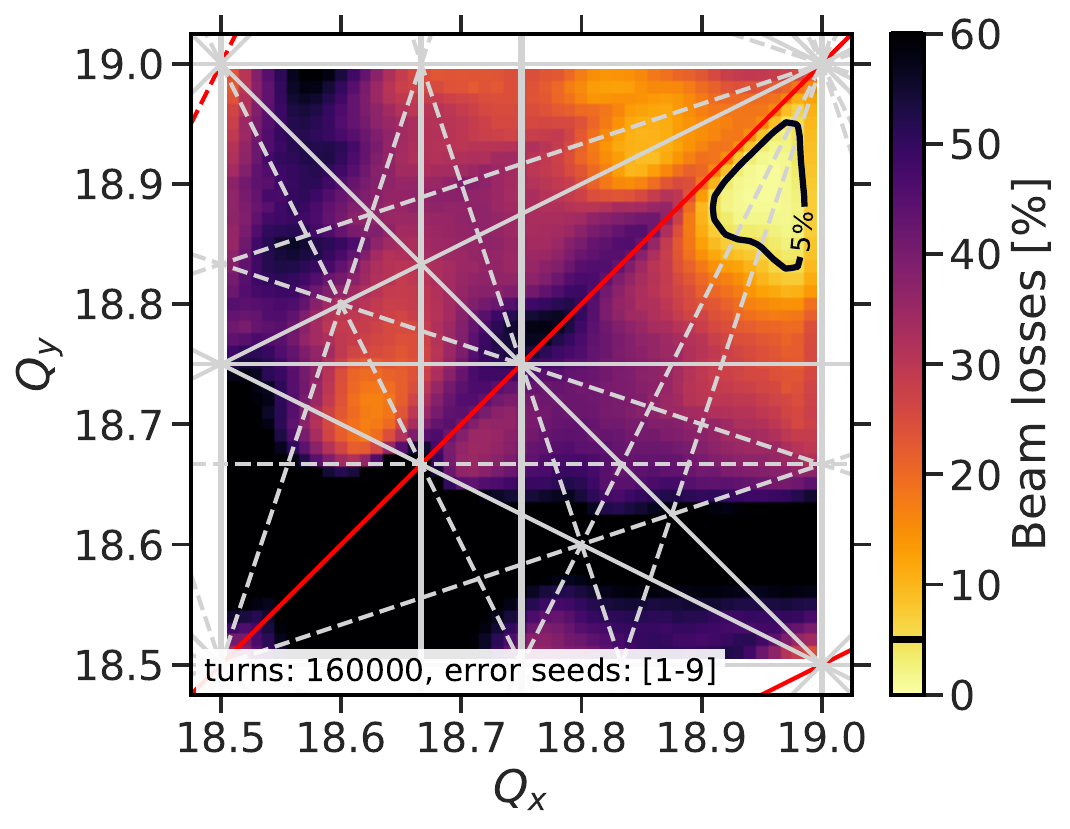}
        \caption{Beam losses.}
        \label{fig: tune diagram nonlinSC allerrs}
    \end{subfigure} \\[1em]
    \begin{subfigure}{\linewidth}
        \includegraphics[width=\linewidth]{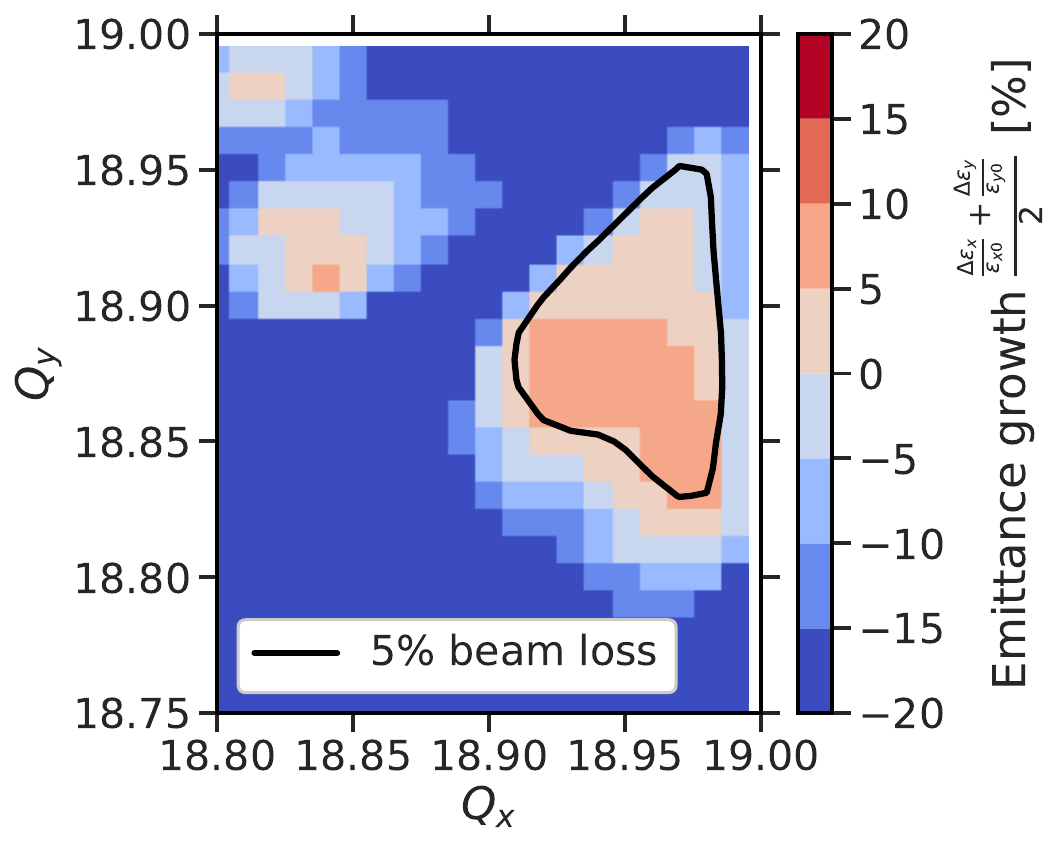}
        \caption{Transverse emittance growth.}
        \label{fig: tune diagram nonlinSC allerrs emittances}
    \end{subfigure}
    \caption{FFSC results for full field error model as a function of the transverse tunes.}
\end{figure}

At first we employ again the FFSC model. The simulation results for the beam losses during the \SI{1}{\second} injection plateau are shown in Fig.~\ref{fig: tune diagram nonlinSC allerrs}.

The central outcome of the FFSC simulation scans at the nominal space charge strength of $\Delta Q_y^\text{SC}=-0.3$ is the identification of a low-loss working point area around $Q_x=18.95$, $Q_y=18.87$.
This triangular area is limited by two solid constraints, namely the Montague resonance on top and the horizontal integer resonance $Q_x=19$ to the right.
While not much can be done to overcome these two restrictions, the limitation at the lower end of the low-loss area is given by higher-order resonances reaching out due to space charge detuning, which can in principle be tackled.

Figure~\ref{fig: tune diagram nonlinSC allerrs emittances} shows the vicinity of the identified (black) low-loss working point area and plots the average of horizontal and vertical emittance growth: red colours indicate a net increase and blue colours a net decrease in transverse emittance.
While the strong beam losses within stop-bands often result in a net shrinking of transverse emittance, the situation looks favourable inside the low-loss area contour:
the FFSC simulations predict an average transverse emittance growth of max.\ $10\%$.
To be precise, the growths amount to some horizontal $\Delta \epsilon_x/\epsilon_x<5\%$ and vertical $\Delta \epsilon_y/\epsilon_y<15\%$.

It is worth to mention that operating SIS100 in this identified working point area below integer tunes requires further considerations.
Firstly, the growth time of the transverse resistive-wall instability becomes short compared to the accumulation time and should be mitigated. Foreseen are passive measures, like dedicated octupole magnets for Landau damping, and an active transverse feedback system.
Appendix~\ref{app: tunes above integer} shows why the adjacent tune quadrants above integer tunes are less favourable for high-intensity operation.
Secondly, the divergence of the closed orbit distortion (which is independent of space charge and originates from magnet misalignments and dipole errors) requires precise control by making use of the steering corrector magnets.
Here it shall suffice to note that corresponding equipment is going to be installed in SIS100.
A detailed discussion of affiliated effects goes beyond the purpose of the present paper and we focus on the space charge aspects.

\begin{figure}[htbp]
    \centering
    \includegraphics[width=0.85\linewidth]{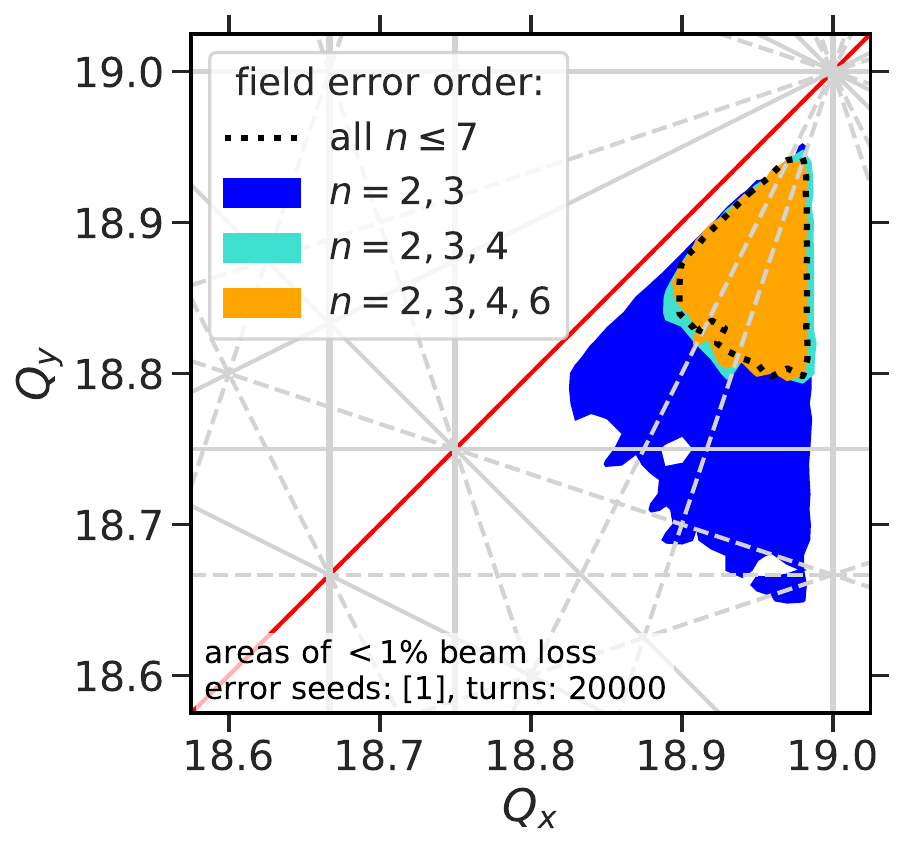}
    \caption{Loss-free tune areas for various included field error orders $n$ in both dipole and quadrupole magnets (FFSC simulations for 20'000 turns).}
    \label{fig: influence error order}
\end{figure}

In order to identify the relevant error multipole orders which limit the low-loss area from below, we performed simulations with reduced error models: the $a_n,b_n$ multipole components in both dipole and quadrupole magnets are cumulatively added with increasing order $n$.
Figure \ref{fig: influence error order} shows the results for tunes below the coupling line, where the coloured areas cover the tune regions of acceptable low-loss working points.
As these simulations only cover 20'000 turns (1/8 of the injection plateau), a threshold of $1\%$ beam loss is chosen.
For reference, the dashed black line represents the low-loss working point area when including all considered field error orders up to 7${}^\text{th}$ order: this contour matches the corresponding $5\%$ beam loss contour after 160'000 turns from Fig.~\ref{fig: tune diagram nonlinSC allerrs}).

The results in Fig.~\ref{fig: influence error order} illustrate the dominating impact of the low-order field errors.
The blue region is limited by resonances excited by the field imperfections of quadrupole and sextupole order.
The turquoise area includes also the octupole order: note the near congruence with the dashed reference contour which includes all errors.
Skipping the decapole order, the orange area further includes the dodecapole order in comparison to the turquoise area, which finally matches the reference dashed black line.
Clearly, the dodecapole order effect is finite but marginal.

Out of the individual $n=5,6,7$ multipole error orders, the dodecapole 6${}^\text{th}$ order is expected to contribute most as concluded from its significant stochastic amplitude (cf.\ Appendix \ref{app: error model}, Fig.\ \ref{fig: quadrupole errors}).
Separate simulations have been carried out considering a limited set of field errors of order $n=2$ plus one out of $n=5,6,7$ (for both dipole and quadrupole magnets).
One indeed finds minor beam loss inflicted through nonlinear resonances only for the $n=6$ case -- the orders of $n=5,7$ have no significant impact at all.
The resonances driven by $n=6$ have negligible influence on the rms beam sizes as opposed to the $n=3$ and $n=4$ cases.

Therefore -- given the field error model for SIS100 --, the quadrupole, sextupole and octupole order and the affiliated resonances together with space charge essentially define the low-loss area.

\subsection{Validation with PIC Simulations}

The key results of Fig.~\ref{fig: tune diagram nonlinSC allerrs} need to be confirmed with PIC simulations of SIS100, now including the full field error model.
The simulations cover a duration of 20'000 turns which is 8 times shorter than the entire \SI{1}{\second} injection plateau.
For this simulation length, a convergence scan using the lattice with self-consistent space charge, but without imperfections, indicates that $10^7$ macro-particles are necessary to sufficiently suppress the PIC-inherent diffusion due to noise and reproduce the expected zero beam loss for a reference working point below the coupling line.

\begin{figure*}[htbp]
    \begin{subfigure}{0.47\linewidth}
        \includegraphics[width=\linewidth]{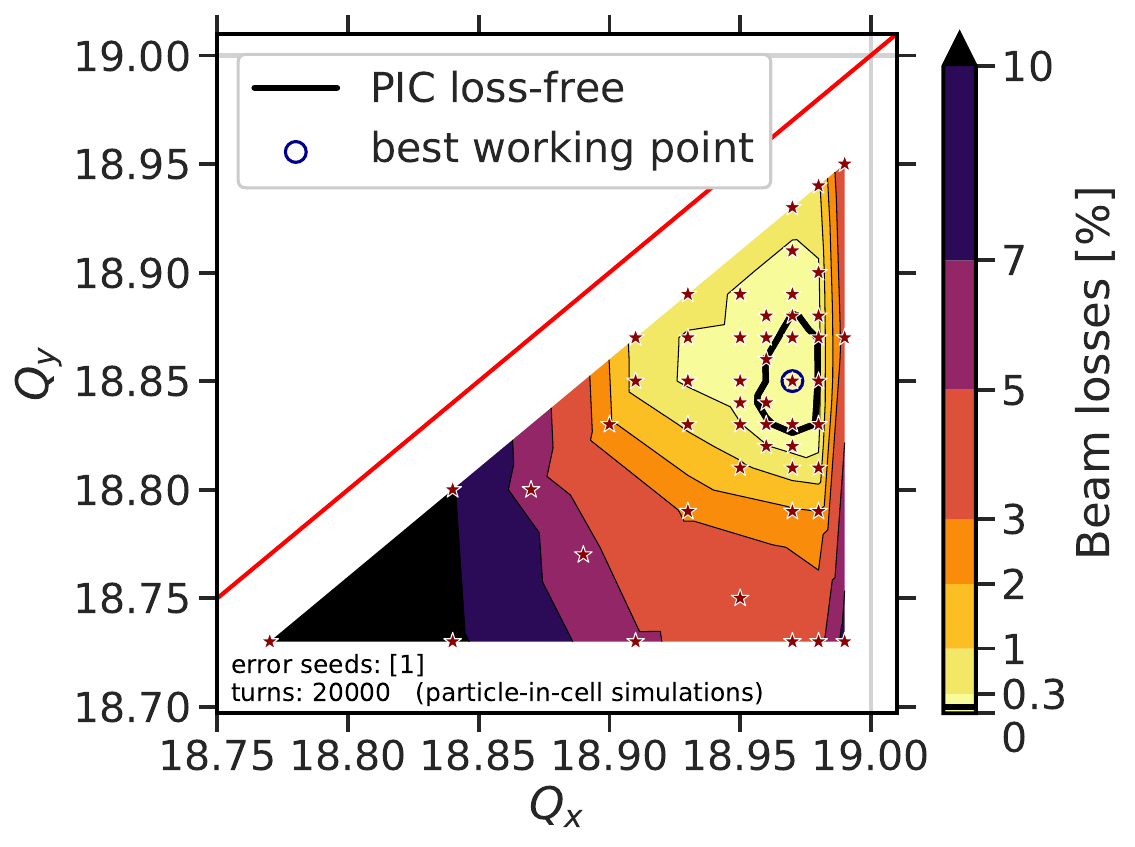}
        \caption{Self-consistent PIC.}
        \label{fig:full-pic}
    \end{subfigure} \hfill 
    \begin{subfigure}{0.47\linewidth}
        \centering
        \includegraphics[width=0.85\linewidth]{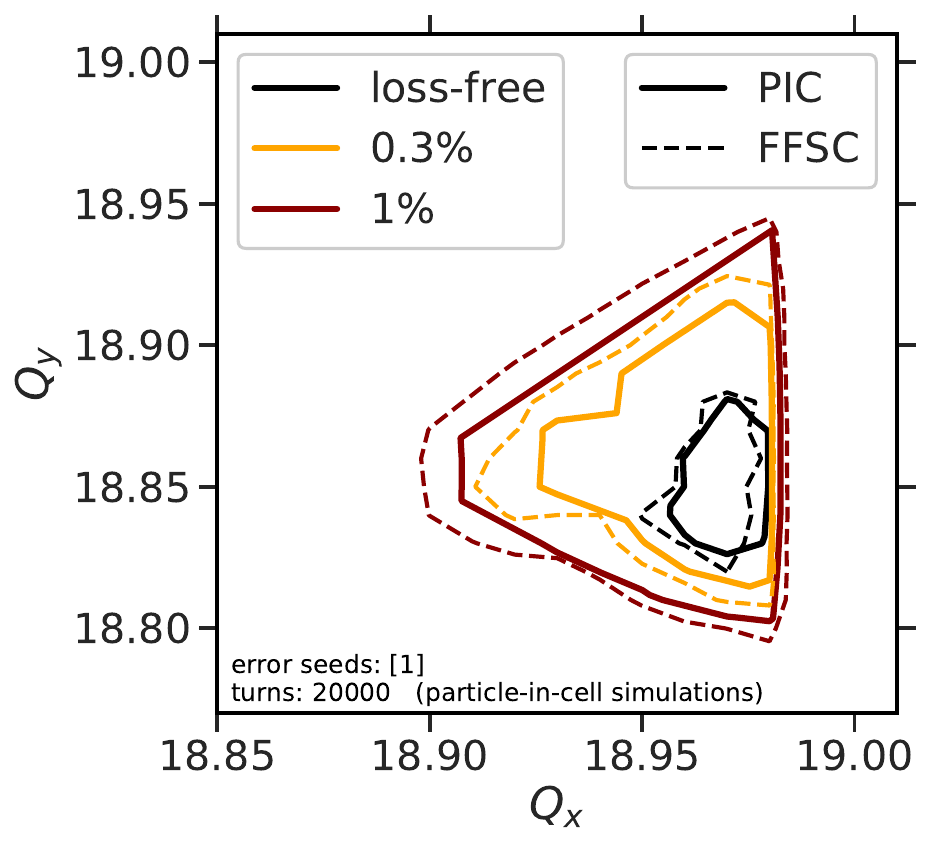}
        \caption{Comparison of predicted loss contours for PIC and FFSC.}
        \label{fig:pic-ffsc-comparison}
    \end{subfigure}
    \caption{Simulation results for beam loss after 20'000 turns using full field error model.}
\end{figure*}

The results of the 2.5D PIC simulations with the full magnet error model are presented in Fig.~\ref{fig:full-pic}.
As opposed to the regular 2D grid of scanned working points in the FFSC simulations before, the more demanding PIC simulations are run only for selected working points indicated by star markers.
The PIC $0.1\%$ beam loss contour is plotted in black: a margin of $0.1\%$ accounts for an upper bound below which PIC-noise-induced losses are difficult to distinguish from resonance losses for the given numerical resolution.
The black contour is thus considered as a ``finite beam loss'' criterion.
The 2.5D approximation of the employed PIC algorithm has been validated by simulating selected working points with a full 3D Poisson solver, which reproduced the 2.5D results.

For direct comparison with the predictions of the FFSC model, Fig.~\ref{fig:pic-ffsc-comparison} shows the contours for areas practically free of loss (black), $0.3\%$ beam loss (orange) and $1\%$ beam loss (dark red) after 20'000 turns.
The solid lines represent PIC results and the dashed lines reproduce the FFSC results.
The dashed black FFSC finite beam loss contour is computed at $0.05\%$ (limit for the numerical FFSC resolution).
Indeed, we find FFSC to predict the same loss-free area as compared to PIC -- in line with the expectations anticipated in Section~\ref{sec:quad-resonances}.

\begin{figure}[htbp]
    \centering
    \includegraphics[width=\linewidth]{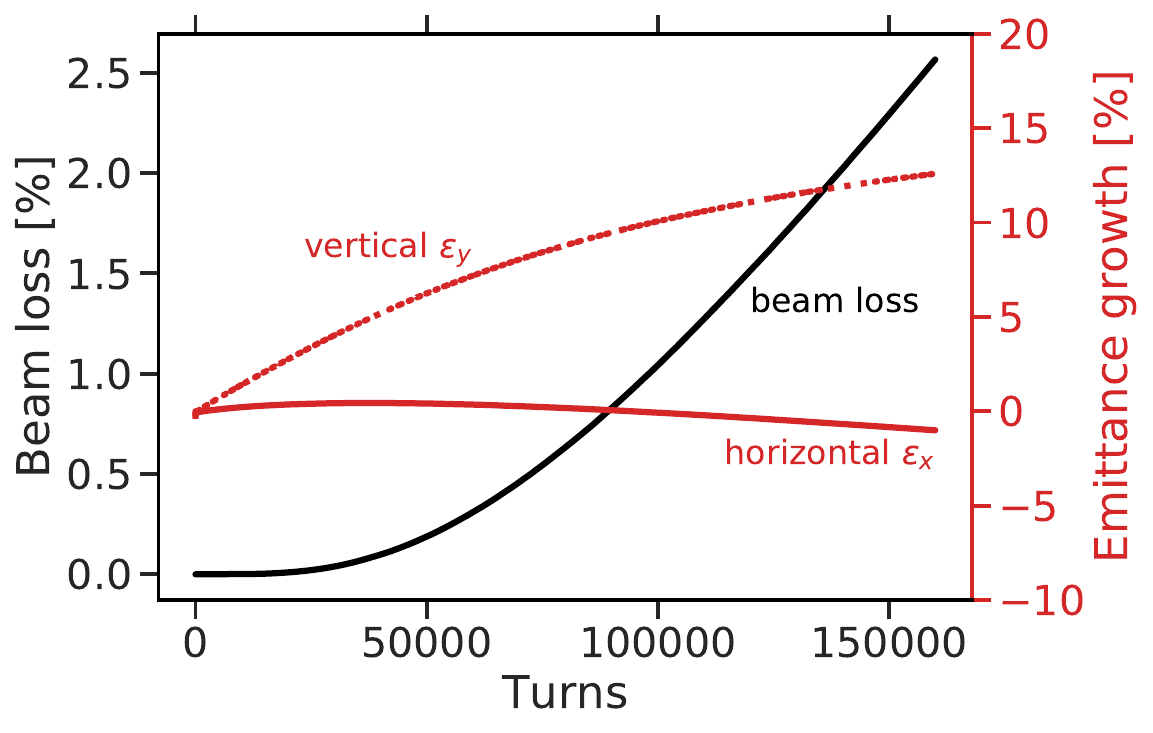}
    \caption{High-resolution PIC results for full field error model at working point $Q_x=18.97$, $Q_y=18.85$.}
    \label{fig:long-pic-sim}
\end{figure}

The working point marked with a blue circle at $Q_x=18.97$, $Q_y=18.85$ exhibits the lowest beam loss overall in the PIC simulations.
We repeat another simulation run for this working point with a further increased resolution ($2\times 10^7$ macro-particles) in order to cover the full \SI{1}{\second} injection plateau.
Figure~\ref{fig:long-pic-sim} shows the beam loss curve in black and emittance growth curves in red.

During the entire storage time of 160'000 turns, the beam is subject to only $2.5\%$ beam loss and a vertical rms emittance growth of $13\%$, the horizontal emittance remains approximately constant.
We conclude that the low-loss working point region identified with the aid of the FFSC model is confirmed by realistic PIC simulations.

To summarise, high-current operation with the envisaged beam parameters of Table~\ref{tab: params} is indeed feasible at strong space charge, achieving $>95\%$ transmission at injection if (a) the magnet imperfections are equal or better than the considered field error model of Figs.~\ref{fig: dipole errors} and \ref{fig: quadrupole errors} and (b) SIS100 is operated on a working point in the vicinity of $Q_x=18.95$, $Q_y=18.87$ for heavy-ion beam production.

\subsection{Space Charge Limit}
\label{sec:sc limit}

At injection energy, beam loss and emittance growth in synchrotrons are often dominated by the space-charge-induced crossing of linear and nonlinear error resonances. 
Assuming that structure resonances can be avoided and only the Montague resonances remains, like in SIS100, a transverse space charge limit can be defined as the maximum bunch intensity corresponding to a given tolerable beam loss or emittance growth. 
Usually the limit is given in terms of the (incoherent) space charge tune shift, corresponding to this maximum intensity.  
Depending on the circumference and the specific conditions the space charge limit will differ for different synchrotrons. 
In the literature on can find more simplistic space charge limits for synchrotrons, such as a frequently quoted maximum incoherent space charge tune shift of $-\Delta Q^\text{SC}\lesssim 0.25$ (see discussion in Ref.~\cite{Weng1987} and references therein). 
Experience from existing synchrotrons shows that the maximum space charge tune shift is not a hard number and is, to date, not possible to accurately predict based only on theory or simulations. 
One reason for this shortcoming can be the incomplete information on the underlying magnet error model, especially the nonlinear error resonances. 
Also the degree of coherent response with space charge can play a role.

\begin{figure}[htbp]
    \centering
    \includegraphics[width=0.9\linewidth]{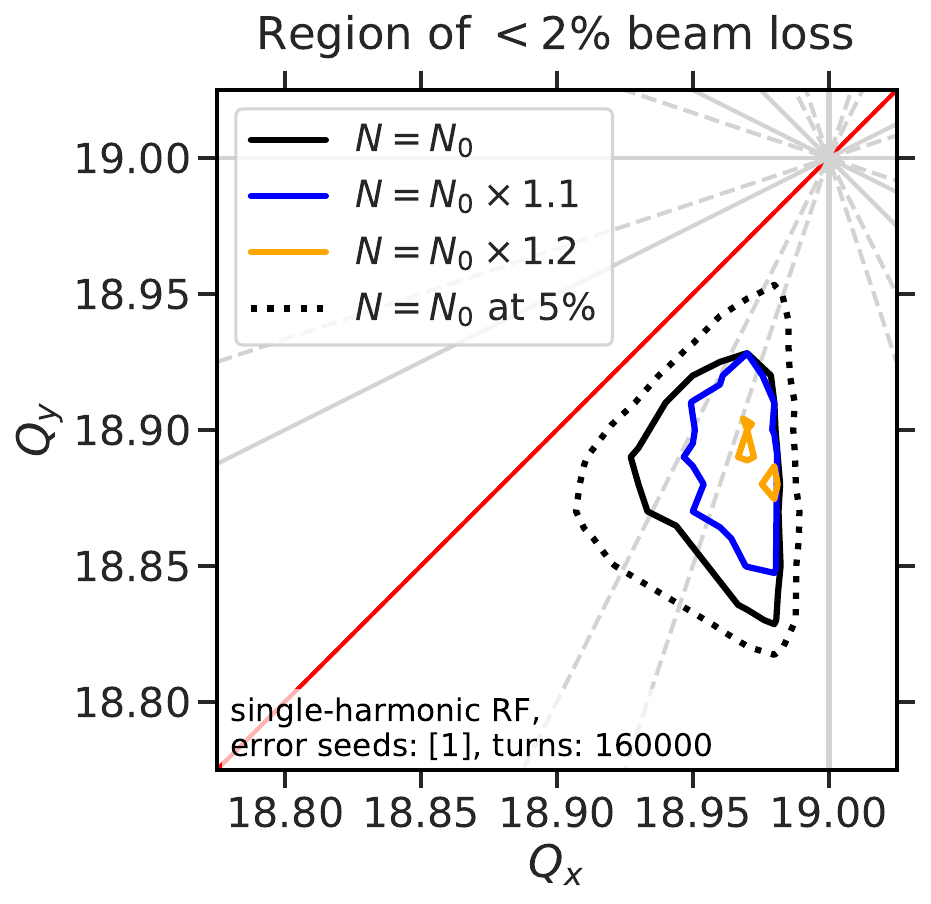}
    \caption{Space charge limit for nominal operation. The solid contour lines enclose the tune area with less than $2\%$ beam loss after the injection plateau for various beam intensities, obtained from FFSC simulations with the full field error model. The dotted contour line shows the reference $5\%$ beam loss threshold as in Fig.\ \ref{fig: tune diagram nonlinSC allerrs}.}
    \label{fig: sc limit h1}
\end{figure}

For the SIS100 we attempt to predict a space charge limit, based on the now existing, detailed magnet error model and for the reference beam parameters.
Only the number of ions per bunch is considered as a free parameter. 
We will employ the FFSC model to estimate the maximum tolerable intensity (all other beam parameters like bunch length, transverse emittances etc.\ are kept constant), for which working points with low beam loss can still be found in the tune diagram. 
For the prediction of the space charge limit we use a slightly stricter beam loss threshold of $2\%$ for the \SI{1}{\second} long injection plateau. This beam loss threshold is motivated by the fact that the corresponding area in tune space (see e.g.\ the solid and dashed black lines in Fig.~\ref{fig: sc limit h1}) is only slightly smaller, but the emittance growth remains substantially lower.
The FFSC simulations use the SIS100 lattice with the full magnet error model.

The same tune scan as for nominal intensity in Fig.~\ref{fig: tune diagram nonlinSC allerrs} has been repeated in simulations with higher intensities, thus gradually increasing the strength of space charge. 
For each intensity setting, contours have been extracted which enclose all simulated working points with a predicted beam loss of $\Delta N/N < 2\%$. 
Figure~\ref{fig: sc limit h1} displays these $2\%$ contours as solid lines for three different intensities. 
Additionally, the $5\%$ contour for the nominal intensity $N_0$ has been plotted in dashed for reference, matching the contour plotted in Fig.~\ref{fig: tune diagram nonlinSC allerrs}.

The available tune space area practically vanishes at $120\%$ of the nominal bunch intensity $N_0$ (orange contours), which corresponds to a maximum space charge tune shift of $\Delta Q_y^\text{SC}=-0.3\times 1.2 = -0.36$.
This intensity marks the space charge limit of SIS100 under nominal operation conditions, above which no low-loss working point is encountered across the considered tune quadrant.
In other words, the maximum achievable space charge tune shift for a conservative beam loss of $2\%$ would be $\Delta Q_y^\text{SC}=-0.36$.

We conclude this section with the remark that, according to our present simulation model, the space charge limit would be slightly above the present reference intensities.

\section{Mitigation Measures}
\label{sec:mitigations}

Here we give three examples of measures to potentially further enlarge the low-loss tune area for the reference beam parameters and to increase the space charge limit.  

\subsection{Correction of $\beta$-beat}
\label{sec:betabeat correction}

Outside the quadrupolar stop-bands the $\beta$-beat is $\approx 0.5\%$ due to the considered 5 units of stochastically distributed gradient errors. This includes feed-down from higher orders in the field error model. Without the local two corrector magnets, the warm quadrupole magnets would be the dominant source of $\beta$-beating ($2\%$).

\begin{figure}[htbp]
    \centering
    \includegraphics[width=0.9\linewidth]{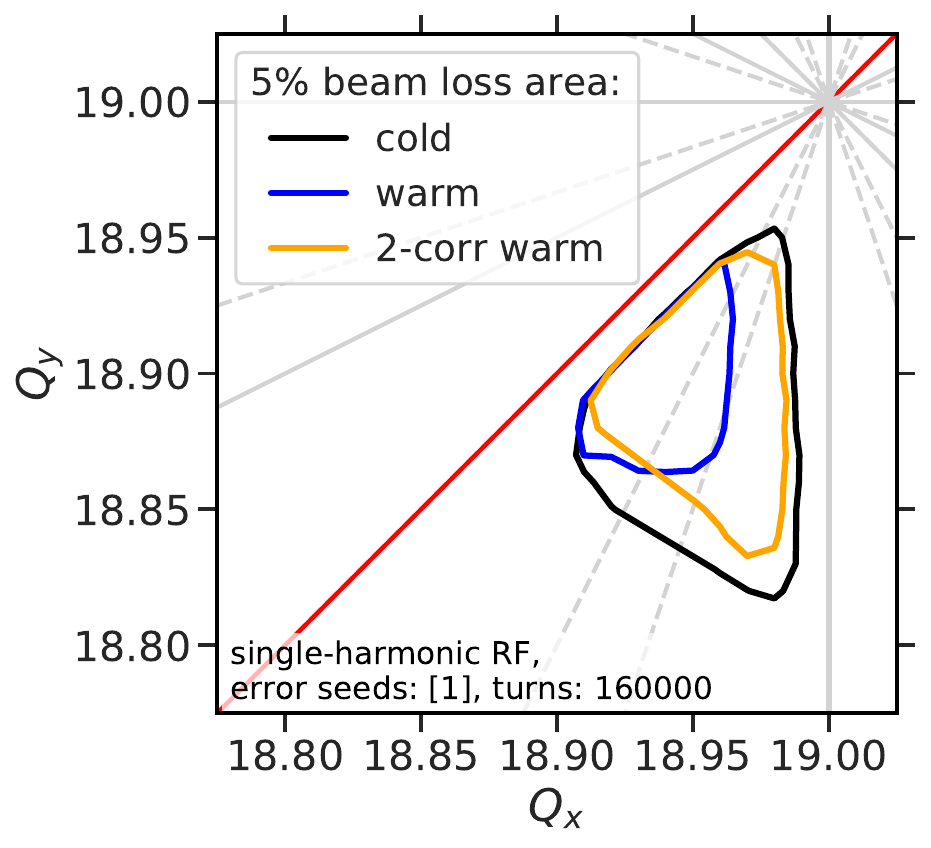}
    \caption{Low-loss area contours for the \emph{cold} symmetric lattice, the broken-symmetry lattice with the two \emph{warm} quadrupole magnets, and the latter plus two corrector magnets. The full injection plateau is simulated with the magnet field error model and the FFSC model.}
    \label{fig: cold-warm-corr}
\end{figure}

Figure \ref{fig: cold-warm-corr} compares the $5\%$ loss contours for the three relevant cases: the symmetric lattice with only \emph{cold} quadrupole magnets used in the beam loss studies (in black), the broken-symmetry lattice with the two \emph{warm} quadrupole magnets (in blue), and the planned operational scenario (in orange), i.e.\ the broken-symmetry lattice including the warm quadrupole magnets and two local correctors.
The orange working point region nearly recovers the fully symmetric case in black.

The impact of the warm quadrupole magnets on the size of the low-loss area is considerable. 
Therefore the local correction scheme with the two corrector magnets suppressing the corresponding $\beta$-beat is essential.

On the other hand the randomly distributed gradient errors in all magnets, the $b_2$, are comparatively weak.
Reduction of the $\beta$-beat induced by $b_2$ from the dipole and quadrupole magnets is not observed to further improve the $<5\%$ low-loss area:
simulations for the ``cold'' lattice yield no significant increase of the loss contours for perfect $\beta$-beat compensation by removing all $b_2$ components.

\begin{figure}[htbp]
    \centering
    \includegraphics[width=\linewidth]{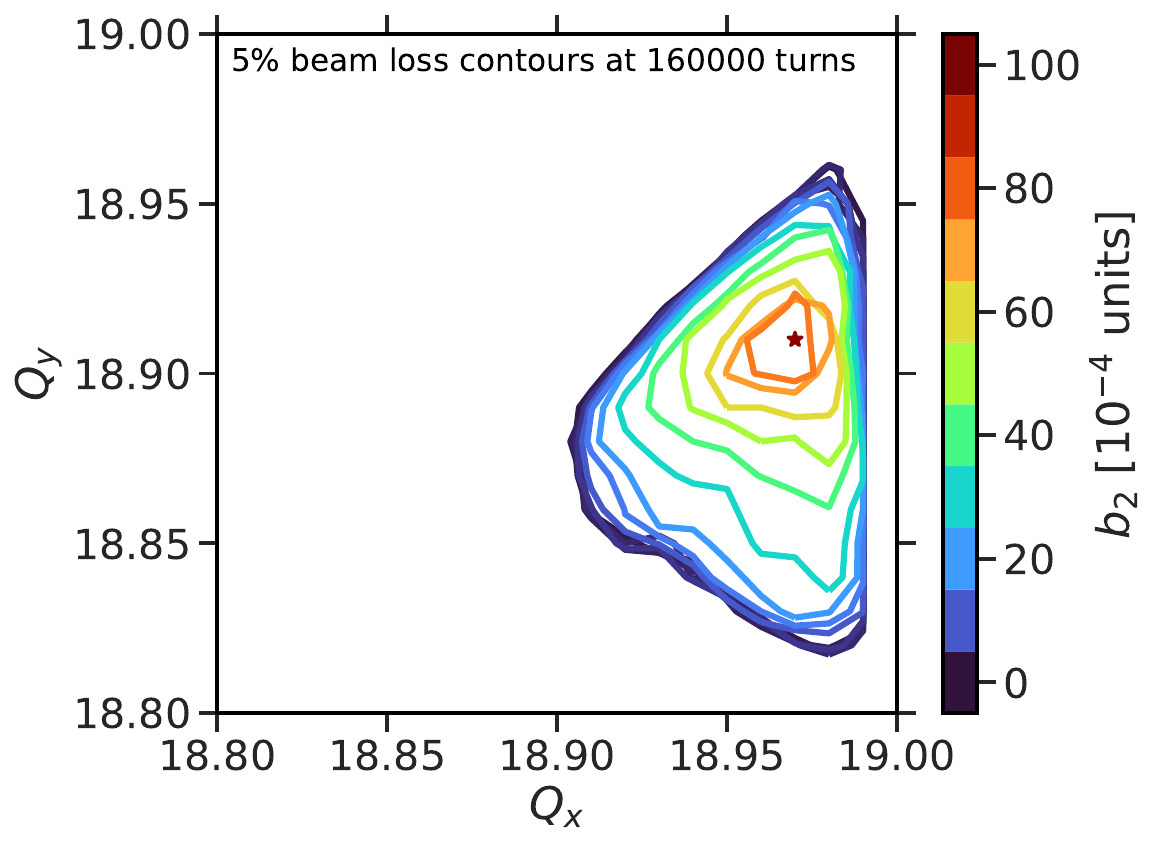}
    \caption{Tune area contours for a $5\%$ beam loss threshold during the injection plateau, plotted for various values of the gradient error $b_2$. The dark red star marks the asymptotic working point around $Q_x=18.97$, $Q_y=18.91$.}
    \label{fig: 5 percent boundaries}
\end{figure}

The impact of increased $b_2$ is best illustrated via a series of tune quadrant simulation scans for various $b_2$ between 0 and 100 units.
Each 2D working point scan for a given $b_2$ setting has been evaluated on a grid of $0.01$ tune distance.
For each $b_2$ setting, the $5\%$ contour of the low-loss area has been plotted in Fig.~\ref{fig: 5 percent boundaries}.
Here, the colour of the contour indicates the gradient error amplitude $b_2$, where the size of the contour shrinks along with the colour gradually moving from dark violet towards yellow.
Eventually the $5\%$ contour vanishes just below $b_2=\SI{80}{units}$.
The corresponding asymptotic working point $Q_x=18.97$, $Q_y=18.91$ is marked by a dark red star.

\begin{figure}[htbp]
    \centering
    \includegraphics[width=\linewidth]{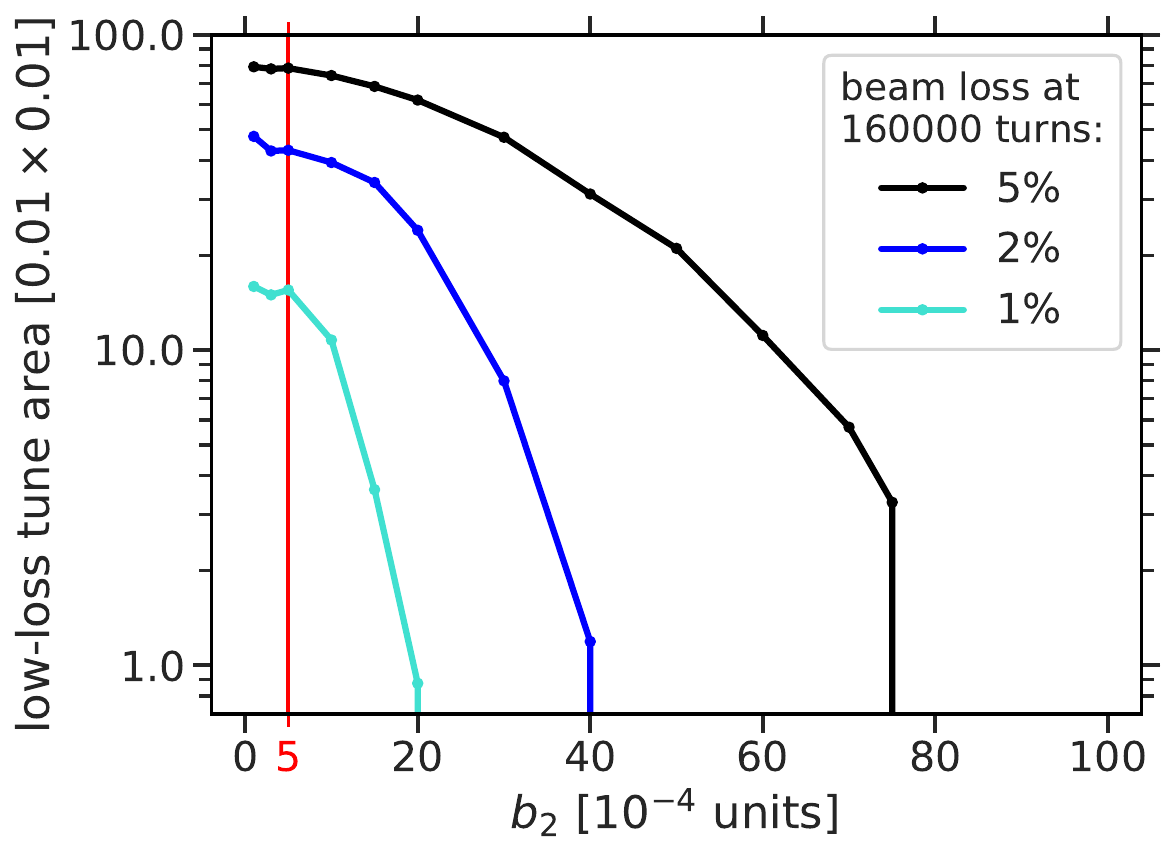}
    \caption{Size of low-loss area as a function of stochastically distributed $b_2$ in the quadrupole magnets (dipole magnets are set to $b_2=0$). The three curves show the $<1\%,<2\%$ and $<5\%$ beam loss areas.}
    \label{fig: b2 scan}
\end{figure}

By integrating the contours one obtains the size of the low-loss area.
Figure~\ref{fig: b2 scan} shows the sizes vs.\ $b_2$ for various levels of the beam loss threshold.
The unit of the vertical axis is hence a measure for ``how many working points correspond to a beam loss less than the respective threshold''.

The red line marks the case of $b_2=5$\,units, which corresponds to the realistic expectation for the quadrupole magnets and which is used in the full field error model throughout this paper.
The plot demonstrates the almost vanishing impact of the expected $b_2=5$\,units in comparison to perfect compensation $b_2=0$.


All in all, the correction of $\beta$-beat which is above an rms figure of about $1\%$ is therefore a viable knob for mitigation of resonance impact.
The level of $\beta$-beat estimated from the present magnet error model is already below this level, provided that the perturbation by the symmetry breaking warm quadrupole magnets is fully compensated.

\subsection{Double-harmonic RF Flattened Bunches}

Bunch flattening via double-harmonic RF operation is a well-known approach for lowering the peak current and thus mitigating transverse space charge.
The CERN Proton Synchrotron Booster for instance successfully operates with such double-harmonic RF systems since 1982 \cite{garoby1997longitudinal}.
In the following we demonstrate the usefulness of this concept for the SIS100.
The ferrite loaded RF cavities in SIS100 offer up to \SI{280}{\kilo\volt} within a frequency window of \SI{1.1}{\mega\hertz} to \SI{3.2}{\mega\hertz} \cite{rfsystems}.
While single-harmonic operation corresponds to an RF frequency of $f_\text{RF}=10f_\text{rev}=\SI{1.57}{\mega\hertz}$, double-harmonic operation during the injection plateau would also fit into the available frequency window at $f_\text{RF}=\SI{3.14}{\mega\hertz}$.

\begin{figure}[htbp]
    \centering
    \includegraphics[width=\linewidth]{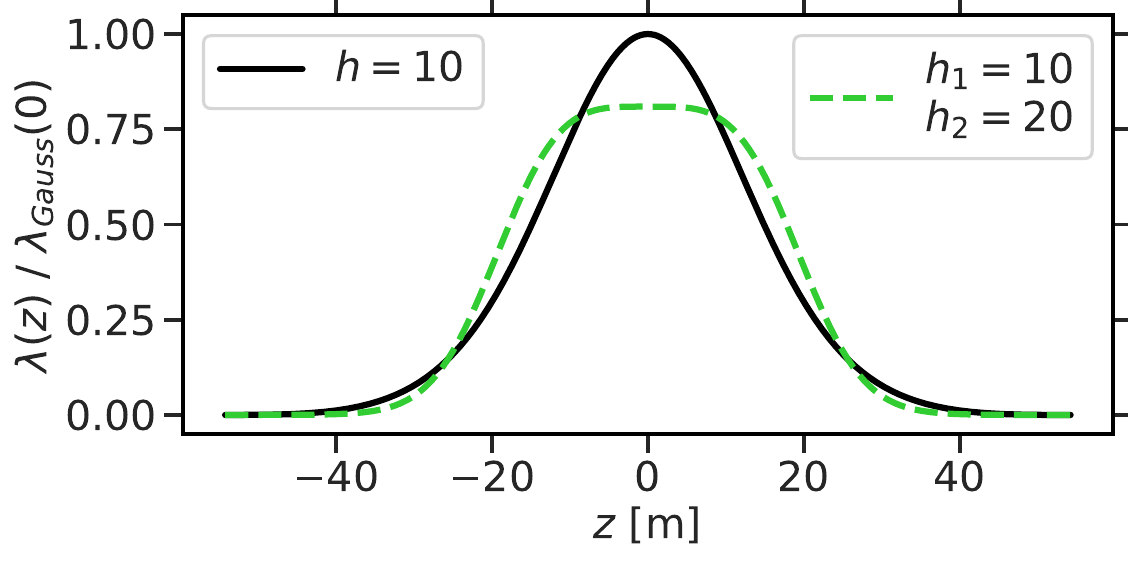}
    \caption{RMS-equivalent bunch profiles for single-harmonic $h=10$ vs.\ double-harmonic $h_1=10$, $h_2=20$ RF operation.}
    \label{fig:doubleharmonic profiles}
\end{figure}

For a better comparison to nominal single-harmonic $h=10$ operation, we choose to keep the rms bunch length $\sigma_z$ and the longitudinal emittance $\epsilon_z$ constant for double-harmonic RF operation.
For bunch lengthening mode, the RF wave of the higher harmonic $h=20$ is phase shifted by $\pi$ in comparison to the $h=10$ RF wave.
The corresponding RF voltage amounts to half the base RF voltage, $V_{h=20}=V_{h=10}/2$.
A value of $V_{h=10}=\SI{103}{\kilo\volt}$ then matches the given longitudinal emittance.
This RF configuration creates a maximally flattened bunch with monotonically decreasing line charge density $\lambda(z)$ from the centre as depicted in Fig.~\ref{fig:doubleharmonic profiles}.
For the chosen bunch flattening mode the transverse space charge force at the bunch centre can thus be reduced by a fifth compared to the rms-equivalent Gaussian bunch profile case.

\begin{figure}[htbp]
    \centering
    \includegraphics[width=\linewidth]{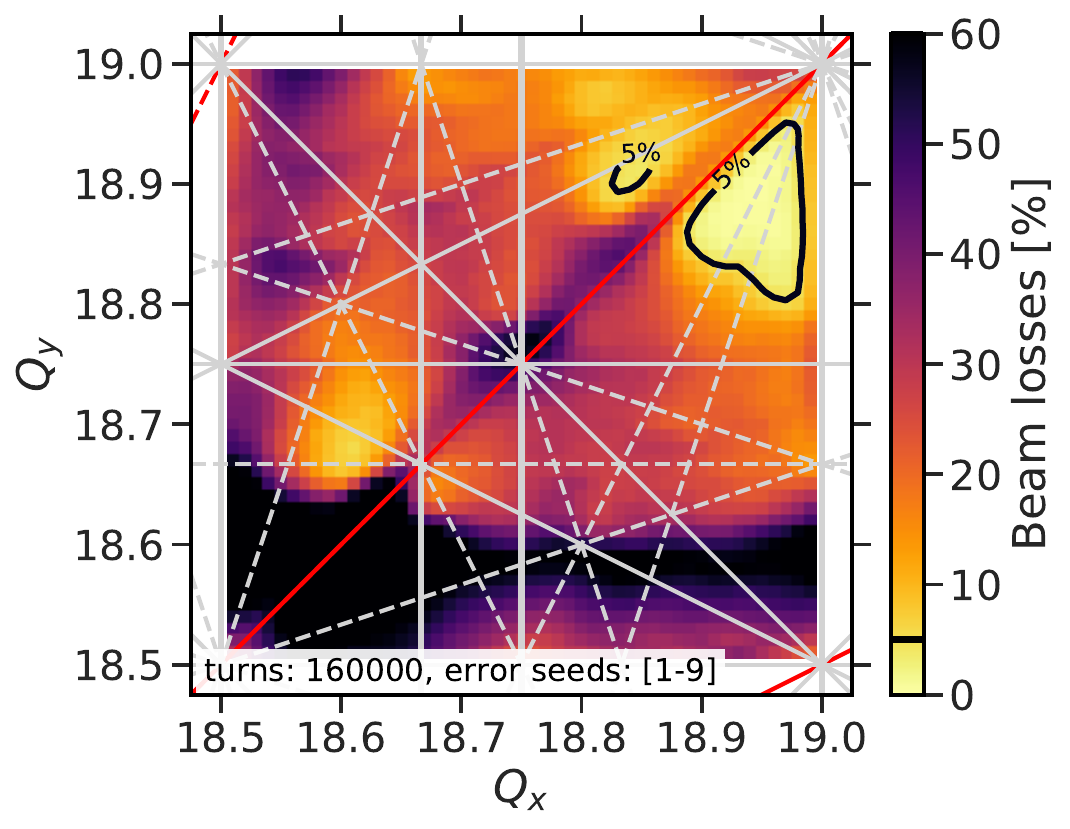}
    \caption{Double harmonic RF operation. Beam loss tune diagram after the injection plateau obtained from the FFSC simulations with the full field error model.}
    \label{fig: tune diagram nonlinSC allerrs doubleharmonic}
\end{figure}

The tune diagram in Fig.~\ref{fig: tune diagram nonlinSC allerrs doubleharmonic} shows the computed beam losses with the FFSC space charge model for the magnet field error model.
The low-loss area extends further in comparison to the equivalent single-harmonic case in Fig.~\ref{fig: tune diagram nonlinSC allerrs}.
The overall magnitude of inflicted beam loss weakens for the entire tune quadrant.
In particular, the upper edge of the black-coloured vertical quadrupole stop-band moves downward towards $Q_y=18.5$ by roughly a fifth, in line with the $20\%$ weaker maximum space charge tune shift.

\begin{figure}[htbp]
    \centering
    \includegraphics[width=0.9\linewidth]{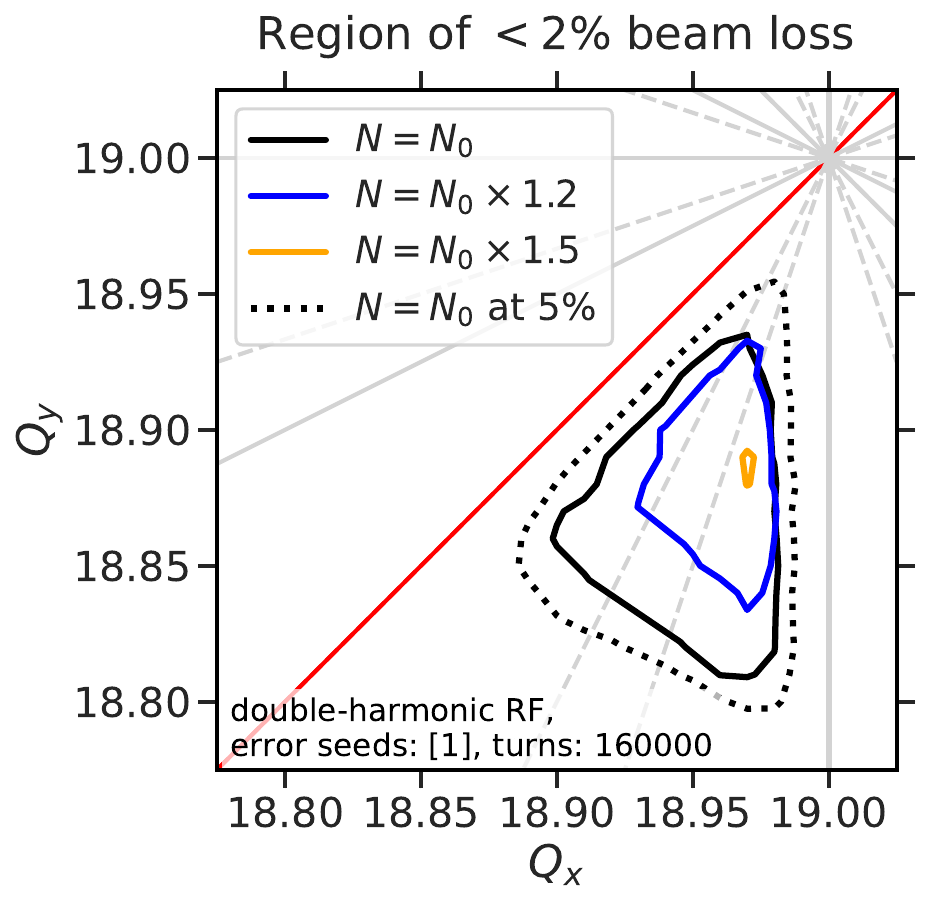}
    \caption{Space charge limit for double-harmonic RF operation. The solid contour lines enclose the tune area with less than $2\%$ beam loss after the injection plateau for various beam intensities, obtained from the FFSC simulation model with the magnet field errors included. The dotted contour line shows the reference $5\%$ beam loss threshold as in Fig.\ \ref{fig: tune diagram nonlinSC allerrs doubleharmonic}.}
    \label{fig: sc limit h2}
\end{figure}

With double-harmonic RF as mitigation measure the SIS100 can accept higher intensities before direct space charge limits the transmission.
Figure~\ref{fig: sc limit h2} shows the results comparing nominal intensity $N_0$ to the single-harmonic RF space charge limit $N_{h_{10}}^\mathrm{max}=1.2N_0$ and the further increased value of $1.5N_0$.
Only in the latter case the available tune space area shrinks to almost zero for the flattened bunches.
Thus, $150\%$ of the nominal intensity injected into the SIS100 marks the space charge limit $N_{h_{10+20}}^\mathrm{max}$ for double-harmonic operation.
We note that the ratio of $150\%$ to $120\%$ in maximally achieved intensity directly relates to the inverse ratio of peak currents.

All in all, we conclude that the traditional space charge mitigation technique of double-harmonic RF bunch lengthening is a useful option for SIS100 heavy-ion operation: more options in tune space are opened up for the reference intensities and the space charge limit can be increased significantly.

\subsection{Pulsed Electron Lenses}

Pulsed electron lenses are currently studied as an optional additional countermeasure for SIS100 to partially compensate the large extent of the space charge tune spread.
The transversely uniform electron beam would be longitudinally shaped to match the passing ion bunch profile.
The opposite charge of the electron lens leads to a partial compensation of the space charge tune spread dependency on the longitudinal position.
The mechanism behind space-charge-induced incoherent scattering and trapping in resonances can thus be inhibited.
The width of resonance stop-bands shrinks and allows for a larger low-loss working point area.
The approach is to install a number of such electron lenses with a transverse homogeneous charge distribution (i.e.\ linear self-fields) in the synchrotron, where the electron pulse moves into the direction of the beam propagation.

A recently published study in Ref.~\cite{elenses-sis100} demonstrates the effect of three of such pulsed electron lenses, which are placed symmetrically around the circumference, in FFSC simulations for SIS100.
The intensity of the electron pulse is adjusted such that the linear part of the space charge force is compensated by half, thus strongly reducing the space charge tune spread.

\begin{figure}[htbp]
    \centering
    \includegraphics[width=\linewidth]{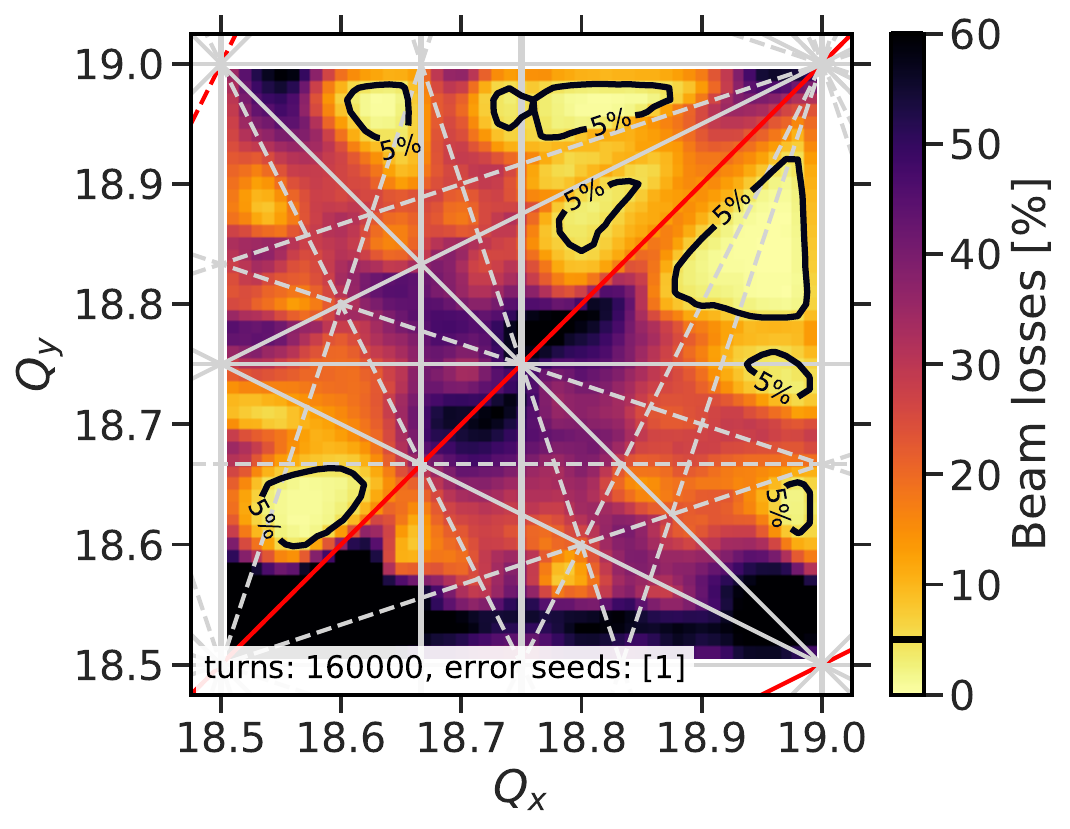}
    \caption{Three pulsed electron lenses mitigate space charge. Beam loss tune diagram after injection plateau, obtained from the FFSC simulation model with the magnet field errors included.}
    \label{fig: elenses}
\end{figure}

For the present study, a set of FFSC simulations has been run for the duration of the entire injection plateau using the parameters of Table \ref{tab: params} (with single-harmonic RF).
The beam loss results with the magnet field errors and the three pulsed electron lenses are presented in Fig.~\ref{fig: elenses}.

In comparison to Fig.~\ref{fig: tune diagram nonlinSC allerrs}, showing the corresponding tune scan without electron lenses, several additional low-loss areas can be identified.
Particularly the width of the quadrupole stop-bands shrinks considerably despite the increased $\beta$-beat implied by the linear optics distortion through the pulsed lenses.
The Montague stop-band appears now symmetrically located around the coupling line $Q_x=Q_y$, shifting down the upper edge of the previously identified low-loss area.
Nonetheless, the low-loss area gains significantly in size as its lower boundary moves downwards to directly above the octupole resonance $4Q_y=75$ as the higher-order resonances are pushed back significantly.

More detailed studies, including tune scans with PIC simulations, are still required. However, the potential of pulsed electron lenses to substantially increase the space charge limit in SIS100 has now been demonstrated.

\section{Conclusion}
This work presents a fast approach to identify suitable working point areas for accumulation of bunched beam in a synchrotron under realistic conditions, where beam loss due to the interaction of magnet-error-induced resonances as well as space charge limits the available area in the tune diagram.
Particular emphasis is given to the systematic comparison of the approximative fixed frozen space charge model (and adaptive variants) with the self-consistent particle-in-cell space charge model, using the FAIR synchrotron SIS100 as an example case.

The framework established in this study turns out to be useful for following up the magnet production quality during series production of the SIS100 quadrupole magnets.
In this way specified tolerances can now be understood and justified from an applied beam dynamics perspective.

To illustrate possible applications, an optimal working point area for SIS100 heavy-ion operation has been identified around $Q_x=18.95$ and $Q_y=18.87$.
Here, bunches featuring space charge tune shifts of up to $\Delta Q_y^\text{SC}=-0.36$ can be stored during the injection accumulation plateau with $>95\%$ transmission.
With the established models at hand, the definition and prediction of the SIS100 space charge limit is carried out.
The results show that the SIS100 design beam parameters feature a margin of a factor $1.2$.
An analysis of the $\beta$-beating induced by stochastic gradient errors illustrates the impact on the space charge limit.

Finally space charge mitigation techniques are evaluated in application to SIS100 and the identified low-loss tune area, determining how they increase the space charge limit: the traditional approach of bunch flattening through double-harmonic RF results in an $\approx 25\%$ increase, while the novel concept of pulsed electron lenses promises a much larger increase.

Further studies into this direction are ongoing, including studies of other tune quadrants foreseen for the different SIS100 operation modes and as possible alternatives for an increased space charge limit.

\begin{acknowledgments}
Carrying out this detailed study was largely enabled based on the fruitful collaboration with the code designers of SixTrackLib, and Adrian Oeftiger wishes to acknowledge the helpful contribution in establishing and accelerating the numerical simulation model by the CERN colleagues Martin Schwinzerl as well as Giovanni Iadarola and Riccardo de Maria.
The authors are also grateful for insightful discussions with the GSI colleagues David Ondreka on SIS100 optics considerations and Kei Sugita on the magnet imperfection modelling.
\end{acknowledgments}

\appendix

\section{Comment on the Adaptive Frozen Space Charge Model}
\label{app: afsc}

The following appendix outlines the issues encountered with the adaptive frozen space charge model (AFSC) when investigating the edges of stop-bands in order to identify resonance-free working points.

\begin{figure}[htbp]
    \centering
    \includegraphics[width=\linewidth]{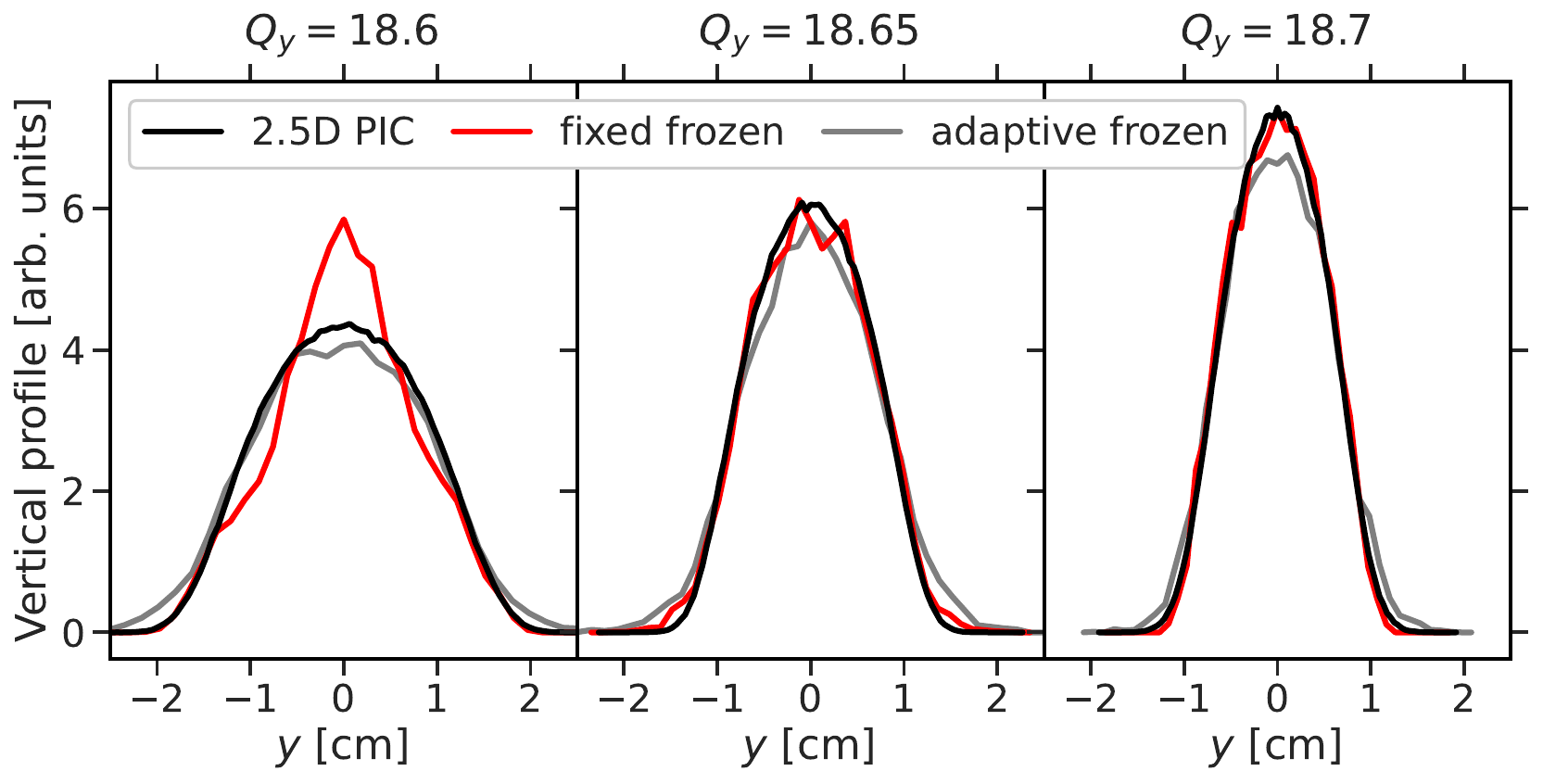}
    \caption{Vertical beam profiles for 3 working points above the $2Q_y=37$ resonance, comparing the PIC results (blue) to FFSC results (orange) after 20'000 turns.}
    \label{fig:profiles}
\end{figure}

As commented in Section \ref{sec:quad-resonances}, the AFSC model provides larger and thus overly conservative rms emittance growth figures towards the resonance-free region in comparison to the PIC and FFSC models.
The difference becomes obvious from the final beam profiles, which are recorded when the resonance effect ceases. 
Figure~\ref{fig:profiles} depicts three panels with beam profiles for three working points from the centre to the upper edge of the stop-band.

\begin{figure*}[htbp]
    \begin{subfigure}{0.47\linewidth}
        \includegraphics[width=\linewidth]{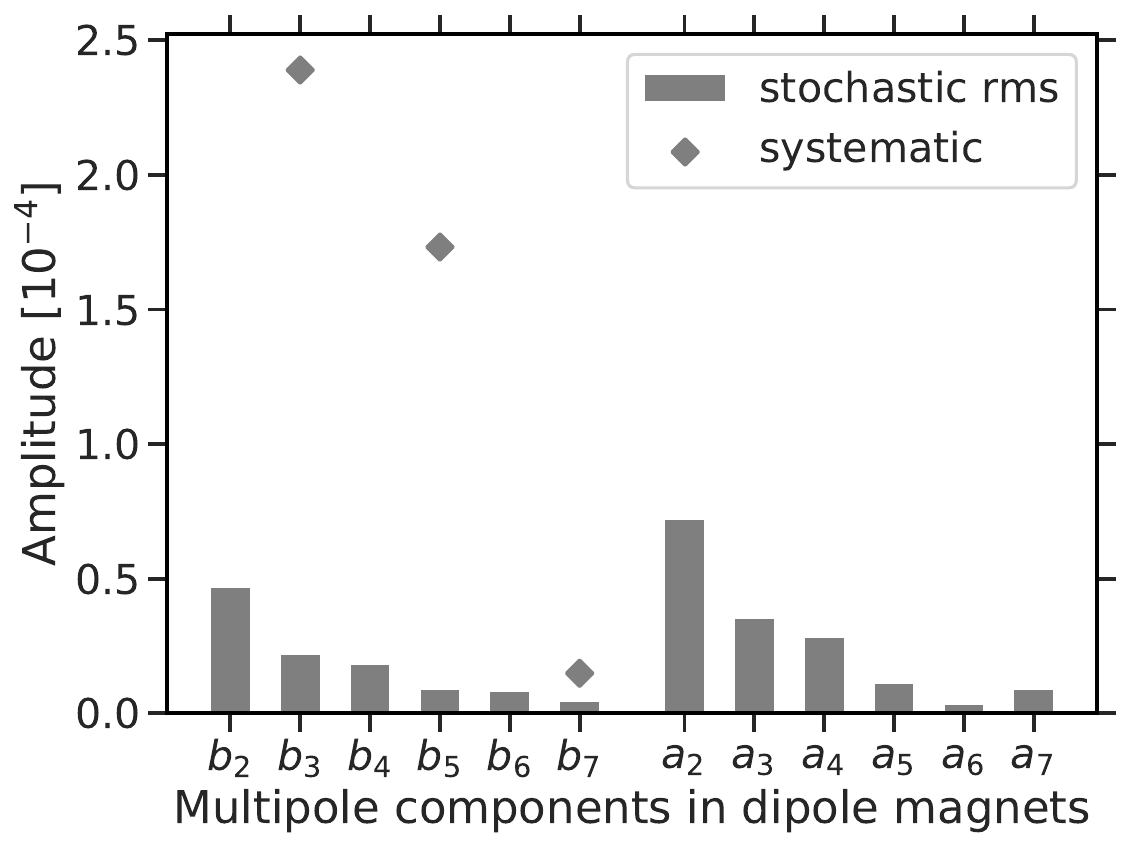}
        \caption{Dipole magnets.}
        \label{fig: dipole errors}
    \end{subfigure} \hfill
    \begin{subfigure}{0.47\linewidth}
        \includegraphics[width=\linewidth]{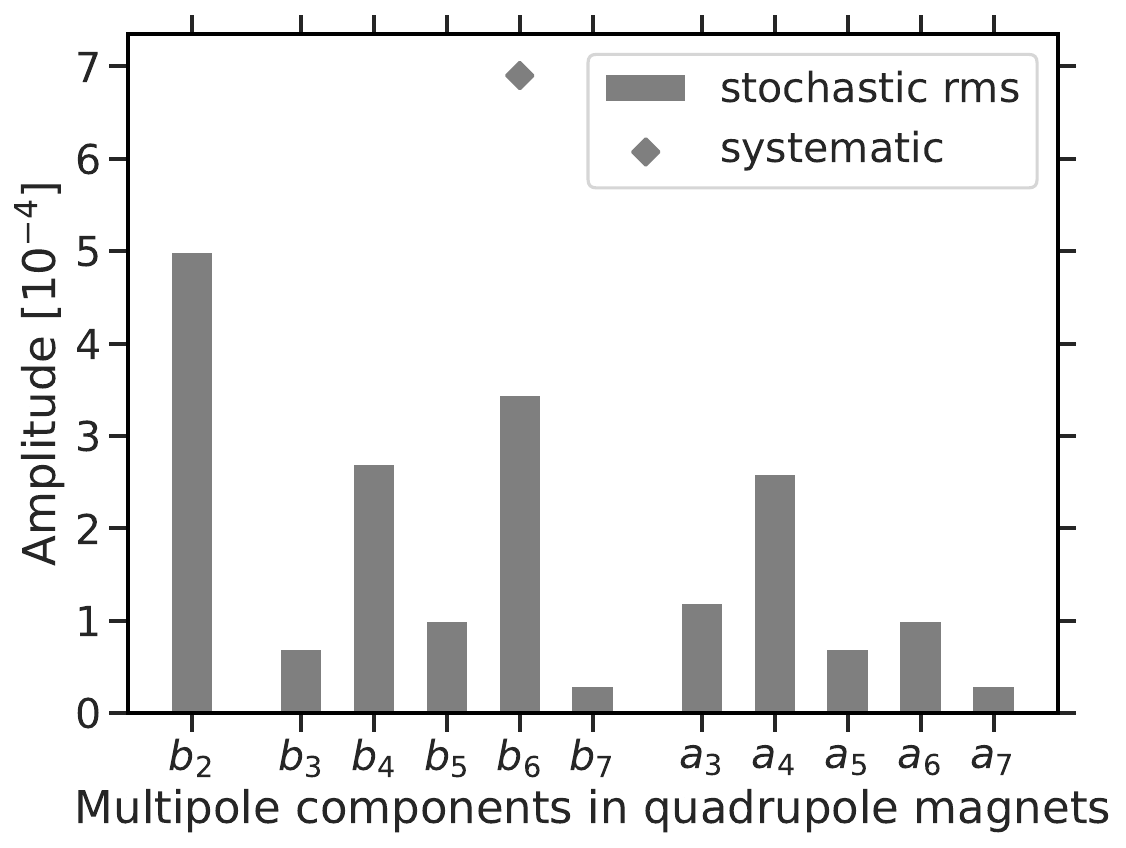}
        \caption{Quadrupole magnets.}
        \label{fig: quadrupole errors}
    \end{subfigure}
    \caption{Field error model based on cold bench measurements.}
\end{figure*}

The left panel shows beam profiles for $Q_y=18.6$ in the centre of the stop-band: they demonstrate how FFSC (red) underestimates the rms emittance growth while PIC (black) and AFSC (grey) predict similar beam profiles.
Towards the upper edge of the stop-band, both the centre panel ($Q_y=18.65$) and the right panel ($Q_y=18.7$) show very similar profiles for FFSC and PIC -- AFSC, however, always exhibits an increased halo population at the profile tails, corresponding to the excess rms emittance growth prediction.

The AFSC model employed here updates the rms beam sizes and centroid offset based on the tracked macro-particle distribution at each space charge node and each turn.
It is worth to note that the emittance growth overshoot is even worse when the centroid is \emph{not} subtracted.

Contrary to the static field maps in FFSC, the adaptive model picks up and amplifies the macro-particle noise through the field updates.
While it is possible to suppress the macro-particle noise impact in the AFSC by choosing larger numbers of macro-particles, relative convergence has only been observed with high numerical resolutions of up to 50 times more macro-particles than FFSC \cite{jinst-sis100-losses}.
Despite the suppressed noise effect, the AFSC results still exhibit a systematic overshoot in the predicted rms emittance growth compared to PIC and FFSC around the upper edge of the stop-band.
A systematic future study could follow up the plausible hypothesis of whether this observation is attributable to the frozen Gaussian nature of the space charge force in conjunction with the increasing rms beam sizes, given the observation that the PIC-simulated distribution shape diverts from a Gaussian.

All in all, the adaptive frozen space charge model provides an overly conservative estimate of resonance-free tune regions compared to the fixed frozen space charge model while at the same time requiring more computational resources.

\section{Field Imperfection Model for SIS100}
\label{app: error model}

In the following, the nonlinear field error model for the SIS100 superferric superconducting dipole and quadrupole magnets is documented as it has been employed in the simulations from Section~\ref{sec: field imperfections} onward.
The development of the error model has been described in detail in Ref.~\cite{jinst-sis100-losses} and is based on cold bench measurements of the real magnets.
Figures~\ref{fig: dipole errors} and \ref{fig: quadrupole errors} exhibit the corresponding multipole components of the field error models for the dipole and quadrupole magnets, respectively.

The bars refer to the rms stochastic amplitudes. 
Magnets around the synchrotron ring in the simulation model are assigned with random values for their respective field imperfection multipoles. 
The random number generator follows a Gaussian distribution which is cut at 2 rms figures of the indicated multipole amplitude.
The diamonds indicate the systematic components, assigned to all magnets as constant offsets in the respective multipole order.

The quadrupole magnets are expected to exhibit a stochastically distributed gradient error of about $b_2=\SI{5}{units}$. 
This value is considered in addition to the nonlinear field errors based on the measurements of the quadrupole magnets.
Section~\ref{sec:betabeat correction} discusses the impact of varying $b_2$, showing that figures beyond $b_2>\SI{10}{units}$ significantly impact the low-loss area.

Misalignment errors in the dipole and quadrupole magnets lead to finite closed orbit distortion (COD) and thus feed-down effects for the magnet imperfections.
For a magnetic field misaligned by $\Delta x, \Delta y$, the contribution of higher-order multipole errors $b_k,a_k$ on lower $b_n,a_n$ is given by
\begin{equation}
    \label{eq: feeddown}
    a_n + i\,b_n = \sum\limits_{k>n}(a_k + i\,b_k) \, \binom{k-1}{k-n} \, \left(\frac{\Delta x + i\,\Delta y}{R_{ref}}\right)^{k-n}
\end{equation}
where $R_{ref}$ denotes the reference radius of the multipole component measurements.
The dipole magnets have been measured at $R_{ref}=\SI{0.03}{\meter}$, the quadrupole magnets at $R_{ref}=\SI{0.04}{\meter}$.
The field error model takes into account an expected rms COD of $x_{co,rms} = \SI{2}{\milli\meter}$ and $y_{co,rms} = \SI{1}{\milli\meter}$ as a global stochastically distributed misalignment in the magnets.
The corresponding feed-down contributions according to Eq.~\eqref{eq: feeddown} are included in the $a_n,b_n$ amplitudes in the simulations.

At present, the series production of the 108 dipole magnets has successfully completed while series production of the 166 superconducting quadrupole magnets is still ongoing.
The currently available measurement data for the first 18 quadrupole magnets shows further reduced imperfections in comparison to the model as shown in Fig.~\ref{fig: quadrupole errors}, cf.\ Ref.~\cite{quadseriesproduction}.
Therefore the simulations presented in this paper represent a conservative prediction for the low-loss tune areas with respect to the expected quadrupole magnet quality in the series production.

\begin{figure*}[htbp]
    \begin{subfigure}{0.47\linewidth}
        \includegraphics[width=\linewidth]{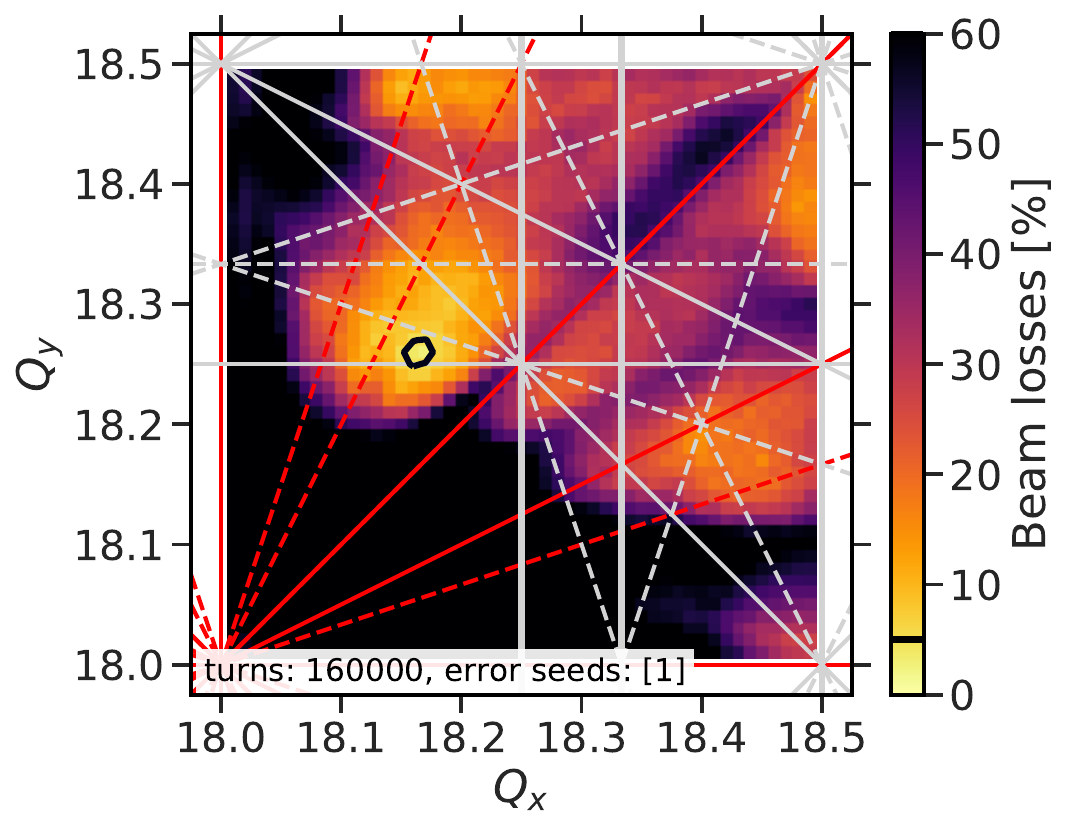}
        \caption{Beam loss above $Q_{x,y}=18$.}
        \label{fig: tune diagram above 18}
    \end{subfigure}
    \hfill
    \begin{subfigure}{0.47\linewidth}
        \includegraphics[width=\linewidth]{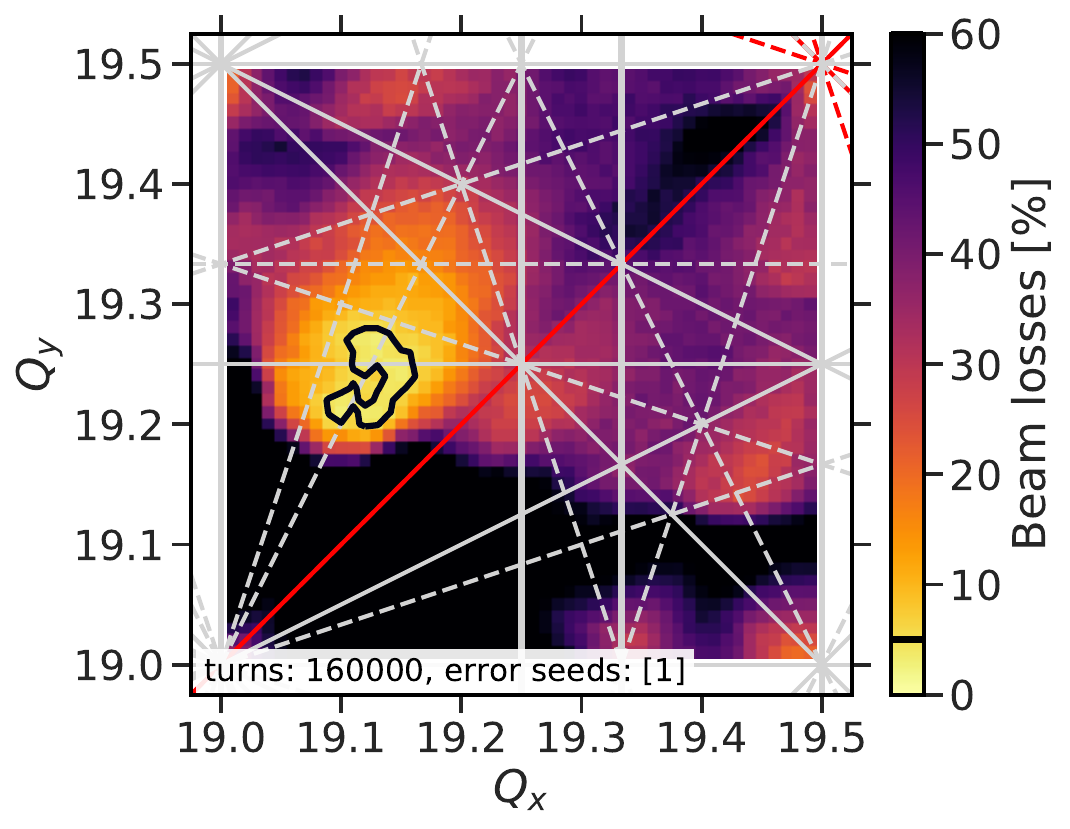}
        \caption{Beam loss above $Q_{x,y}=19$.}
        \label{fig: tune diagram above 19}
    \end{subfigure}
    \caption{FFSC results for the full field error model as a function of the transverse tunes above the integer resonances.}
    \label{fig: tune diagrams above integers}
\end{figure*}

\section{Tune Quadrants above Integers}
\label{app: tunes above integer}

In Ref.~\cite{Franchetti2006b} the tune quadrant $18.5<Q_{x,y}<19$ has been identified as a good candidate for operation with fast extraction of heavy-ion beams. 
Tunes below the integer resonance require the resistive-wall instability to be suppressed, which in case of SIS100 is achieved via the available dedicated octupole magnets for Landau damping and an active transverse feedback system. 
Tunes above the integer resonance are favourable in terms of the resistive-wall instability, which is why most high-intensity synchrotrons are operated in this regime.
In the case of SIS100, however, the nearby located tune quadrants above the integer tunes $18$ and $19$ suffer from increased resonance-induced losses compared to the nominal tune quadrant.
The corresponding loss tune diagrams are displayed in Fig.~\ref{fig: tune diagrams above integers}. 
The black $5\%$ loss contours enclose much smaller low-loss tune areas than in Fig.~\ref{fig: tune diagram nonlinSC allerrs}. 
No working points are found with losses below $1\%$ across the simulated \SI{1}{\second} injection plateau.
The presence of space-charge-induced and low-order structure resonances explain the increased difficulty to find low-loss tune areas in these two tune quadrants above the integer resonances.
We therefore conclude that, given that coherent stability is ensured, the nominal tune quadrant looks favourable for high-intensity operation in SIS100.

\bibliography{bibliography}

\end{document}